\documentclass[journal]{IEEEtran}
\usepackage{amsmath,amsfonts}
\usepackage{amssymb}
\usepackage{amsthm}
\usepackage{algorithm,algpseudocode}
\usepackage{algorithm}
\usepackage{array}
\usepackage[caption=false,font=normalsize,labelfont=sf,textfont=sf]{subfig}
\usepackage{textcomp}
\usepackage{stfloats}
\usepackage{url}
\usepackage{ulem}
\usepackage{verbatim}
\usepackage{graphicx}
\usepackage{xcolor}
\usepackage{ulem}
\usepackage{cite}
\usepackage{booktabs}
\begin{document}

\title{Semantic Entropy Can Simultaneously Benefit Transmission Efficiency and Channel Security of Wireless Semantic Communications} 


\author{Yankai Rong, Guoshun Nan,~\IEEEmembership{~Member,~IEEE,} Minwei Zhang, Sihan Chen, Songtao Wang, Xuefei Zhang, \\Nan Ma,~\IEEEmembership{~Member,~IEEE,} Shixun Gong, Zhaohui Yang, Qimei Cui,~\IEEEmembership{~Senior Member,~IEEE,} \\Xiaofeng Tao,~\IEEEmembership{~Senior Member,~IEEE,}   Tony Q.S. Quek,~\IEEEmembership{~Fellow,~IEEE}
\thanks{Y. Rong, G. Nan, M. Zhang, S. Chen, X. Zhang, N. Ma, S. Gong, Q. Cui, X. Tao are with Beijing University of Posts and Telecommunications, China and National Engineering Research Center for Mobile Network Technologies. (e-mail: muyehu016@gmail.com; nanguo2021@bupt.edu.cn; mmwzhang@gmail.com; chensihanaw@163.com; zhangxuefei@bupt.edu.cn; manan@bupt.edu.cn; sxgongbeiyou@bupt.edu.cn; cuiqimei@bupt.edu.cn; taoxf@bupt.edu.cn).}
\thanks{S. Wang (wangst@zgclab.edu.cn) is an assistant researcher at Zhongguancun Laboratory, Beijing, China.}
\thanks{Z. Yang is with Zhejiang University, Hangzhou, China. (E-mail: yang\_zhaohui@zju.edu.cn).}
\thanks{T. Q. S. Quek is with the Singapore University of Technology and Design, Singapore 487372, and also with the Yonsei Frontier Lab, Yonsei University, South Korea (e-mail: tonyquek@sutd.edu.sg).}
\thanks{Corresponding author: G. Nan (e-mail: nanguo2021@bupt.edu.cn).}}

\markboth{Journal of \LaTeX\ Class Files,~Vol.~14, No.~8, August~2021}%
{Shell \MakeLowercase{\textit{et al.}}: A Sample Article Using IEEEtran.cls for IEEE Journals}

\IEEEpubid{0000--0000/00\$00.00~\copyright~2021 IEEE}

\maketitle

\begin{abstract}
Recently proliferated deep learning-based semantic communications (DLSC) focus on how transmitted symbols efficiently convey a desired meaning to the destination. However, the sensitivity of neural models and the openness of wireless channels cause the DLSC system to be extremely fragile to various malicious attacks. This inspires us to ask a question: ``Can we further exploit the advantages of transmission efficiency in wireless semantic communications while also alleviating its security disadvantages?''. Keeping this in mind, we propose SemEntropy, a novel method that answers the above question by exploring the semantics of data for both adaptive transmission and physical layer encryption. Specifically, we first introduce semantic entropy, which indicates the expectation of various semantic scores regarding the transmission goal of the DLSC. Equipped with such semantic entropy, we can dynamically assign informative semantics to Orthogonal Frequency Division Multiplexing (OFDM) subcarriers with better channel conditions in a fine-grained manner. We also use the entropy to guide semantic key generation to safeguard communications over open wireless channels. By doing so, both transmission efficiency and channel security can be simultaneously improved. Extensive experiments over various benchmarks show the effectiveness of the proposed SemEntropy. We discuss the reason why our proposed method benefits secure transmission of DLSC, and also give some interesting findings, e.g., SemEntropy can keep the semantic accuracy remain 95$\%$ with  60$\%$ less transmission.

\end{abstract}


\section{Introduction}
\IEEEPARstart{W}{ith} the rapid growth of wireless applications and the continuous increase in data traffic, the field of wireless communication is facing a serious challenge in the form of spectrum scarcity. This challenge has prompted a paradigm shift in communication from traditional communication modes towards the more efficient paradigm of semantic communication \cite{all-JSNC,all-speech,all-7,all-8,all-9,all-10,all-11}. DLSC have shown great potential in improving communication efficiency by only delivering the semantics conveyed in transmitted data, in contrast to the traditional Shannon paradigm that aims to guarantee the correct reception of raw bits. The principle of DLSC is to explore the capability of deep learning techniques to transform the data into high-dimensional semantic representations by neural networks at a transmitter, send the encoded symbols to wireless channels, and then reconstruct or interpret the semantics at the receiver side \cite{all-2,all-3,all-4,all-5,all-6}. Hence, such goal-oriented DLSC can significantly facilitate the deployment of mobile applications that require high throughputs and low latency, such as VR/AR and Human-to-Human communications \cite{all-view,all-research,DeepSC-health,meta,car}.

In conventional block-wise communication systems, the modules in the transmitter and receiver are individually designed with different assumptions and objectives, making it difficult to ascertain the global optimality of the system \cite{tra,tra2,tra3}. The novel communication paradigm of DLSC opens up an opportunity for jointly optimizing the transmitter and receiver modules in an end-to-end fashion. Recent efforts focus on designing powerful semantic encoders and decoders for information extraction and reconstruction. There are also a line of works that aim to jointly optimize the semantic encoder and channel encoder \cite{jc,DeepSC,DeepSC-HARQ,GAN,image-ICASSP,federated,DeepSC-health,CVPR1,TIFS2}. However, wireless modulation, which encodes data into a signal that can be sent through various physical media, has yet to be fully considered in the optimization of goal-oriented semantic communications, although previous works have discussed the effectiveness of OFDM for semantic transmission \cite{JSCC-OFDM,JSCC-OFDM2}. Meanwhile, due to the opacity of deep learning models, semantic features extracted from source data are encoded and coupled together. The black-box nature of deep learning operations obstructs the targeted application of specific semantic features, which may lead to unintended semantic features being transmitted to the receiver, reducing the efficiency of data transmission \cite{DeepSC-MT,DeepSC,resource,resource2,resource3,resource4}.
\IEEEpubidadjcol

DLSC system also suffers from more challenging security issues  compared with traditional wireless communication: the sensitivity of neural models and the openness of wireless channels may also cause the DLSC system to be extremely fragile to eavesdropping, tampering, and spoofing \cite{DeepSC-S,Sem-security,ICC,Sem-security2,Sem-security3,Sem-security4,security5,security6,TIFS}. Early efforts in block-wise communication systems use symmetric-key or asymmetric-key encryption schemes, while these methods may lead to heavy computation and communication overhead \cite{security1,security2,security3,security4}. Physical layer key (PLK) schemes reduce the overhead by utilizing the reciprocity principle and temporal variation of the wireless channels between two legitimate users for encryption and decryption \cite{OFDM,OFDM-chaos,chaos-1,chaos-2}. Towards this direction, recent work \cite{ICC} relies on the BLEU scores in natural language processing to generate secret keys to secure text-based semantic communication systems. However, it needs real-time interactions between the semantic encoder and decoder to obtain text-based BLEU scores. Hence, such a method can hardly be applied to other DLSC systems.

The aforementioned discussions inspire us to consider how to exploit the advantages of transmission efficiency in wireless semantic communications while also alleviating its
security disadvantages. To this end, we propose SemEntropy, a novel DLSC system that aims to jointly optimize transmission efficiency and secure communications by a semantic-guided subcarrier allocation and physical layer encryption scheme. Semantic entropy indicates the expectation of various semantic scores regarding the transmission goal.  
Our high-level idea is to rely on such semantic entropy to automatically perform a semantic importance generator on the input of the target-oriented DLSC and select the information to be transmitted based on semantic entropy constraints. Then, we use semantic entropy to guide the encryption based on physical layer secret key generation as well as the joint optimization based on adaptive OFDM subcarrier allocation. 

Specifically, our SemEntropy first learns to deconvolute the input image through the semantic extraction module in the semantic information generator, to distinguish the image into different levels under specific communication targets. We also introduced the importance score of semantic information. Simultaneously, based on the proposed semantic entropy, quantitatively select the transmitted semantic information. Then we present two key components based on semantic entropy, i.e., semantic key generator and adaptive subcarrier allocator. 1) The former relies on the semantic scores to generate semantic secret keys for the encryption of semantic data, respectively, and it explores semantic bias and the reciprocity and temporal variation of the wireless channels to introduce more randomness; 2) The latter utilizes the importance scores to distribute the semantic data to different subcarriers, where the data with higher scores will be assigned to a stronger subcarrier. By doing so, significant semantic information will be transmitted in high priority to facilitate the reconstruction or interpretation at the receiver side. We train the DLSC system with such a novel modulation scheme in an end-to-end manner to jointly optimize communication efficiency. Experiments verify our SemEntropy. We summarize our contributions as follows:

\begin{itemize}
\item We propose SemEntropy, a novel DLSC system that relies on semantic entropy to jointly optimize transmission efficiency with adaptive subcarrier allocation and secure communications with physical layer encryption. 
\item We present two key components, i.e., semantic key generator and adaptive subcarrier allocator, where the former and latter rely on entropy scores to craft secret keys for semantic data, and allocate different subcarriers, respectively.  
\item We conduct extensive experiments to show the effectiveness of this method and further provide quantitative analysis based on search space and bit error rate performance to confirm the security of the encryption.
\end{itemize}

\section{Related Work}
\subsection{Semantic Communication Systems}
In recent years, semantic communication has attracted widespread research interest. Semantic communications transmit semantics conveyed in the data rather than raw bits. Previous works mainly focus on the design of powerful encoders and decoders or joint optimization. 

In the field of text transmission, the transformer-based text transmission system DeepSC has been proposed in \cite{DeepSC}. Its primary goal is to maximize the system capacity and minimize semantic errors by recovering the meaning of sentences. Building upon this foundation, a variant of DeepSC has been developed in \cite{DeepSC-S}.  Furthermore, a hybrid Automatic Repeat reQuest (HARQ) mechanism has been exploited in \cite{DeepSC-HARQ} to further reduce semantic transmission errors. The research on text transmission has been extended to various application scenarios. A transformer-based framework has been proposed in \cite{DeepSC-MT} to unify the structure of transmitters for different tasks. \cite{DeepSC-health} introduces semantic cognitive concepts into the health service system, proposing a health service network framework.

In addition, researchers have also developed semantic communication systems tailored for image and video transmission. \cite{GAN} proposes a Reinforcement Learning-based Adaptive Semantic Coding (RL-ASC) approach that encodes images beyond the pixel level. 
Additionally, \cite{SVC} develops a Semantic Communication System (SVC) for video conferencing and employed HARQ and Channel State Information (CSI) to enhance transmission quality. These studies have expanded the possibilities of semantic communication systems, demonstrating their effectiveness not only in multi-user scenarios \cite{DeepSC-MT} but also in multi-task contexts \cite{muti-task}.

\subsection{OFDM Physical Layer Encryption}
OFDM is a digital communication technology widely used in high-speed data transmission and wireless communication standards. It enhances communication efficiency and reliability by segmenting data streams into multiple orthogonal subcarriers. Research on OFDM systems can be traced back to 1971 \cite{OFDM}. However, due to the open nature of wireless channels, unencrypted OFDM systems are vulnerable to eavesdropping attacks. Physical layer security, with its simplicity and ease of use, offers a viable solution to safeguard the security of OFDM systems. Early research \cite{OFDM-chaos} introduces a chaos-based OFDM constellation scrambling method for physical layer security. Towards this direction, recent efforts aim to utilize scrambling or interleaving techniques to protect OFDM systems\cite{chaos-1,chaos-2}. \cite{chaos-1} divides Quadrature Amplitude Modulation (QAM) symbols into two subsets and transmitted them on encrypted/non-encrypted subcarriers, making it more challenging for eavesdroppers to correctly detect transmitted data. 

However, in static fading environments such as indoor Internet of Things, the temporal decorrelation of the wireless channel is not guaranteed, often leading to significantly reduced key generation rates, approaching ultra-low or even zero. Existing PLK efforts primarily rely on additional relays or Reconfigurable Intelligent Surfaces (RIS) to exploit the randomness injected into the wireless channel within static environments. In contrast to these endeavors, this paper proposes a novel physical-layer semantic encryption/decryption approach by exploring the stochasticity of information processing within semantic communication systems.

\begin{figure*}[htbp]
\centerline{\includegraphics[width=1.0\linewidth]{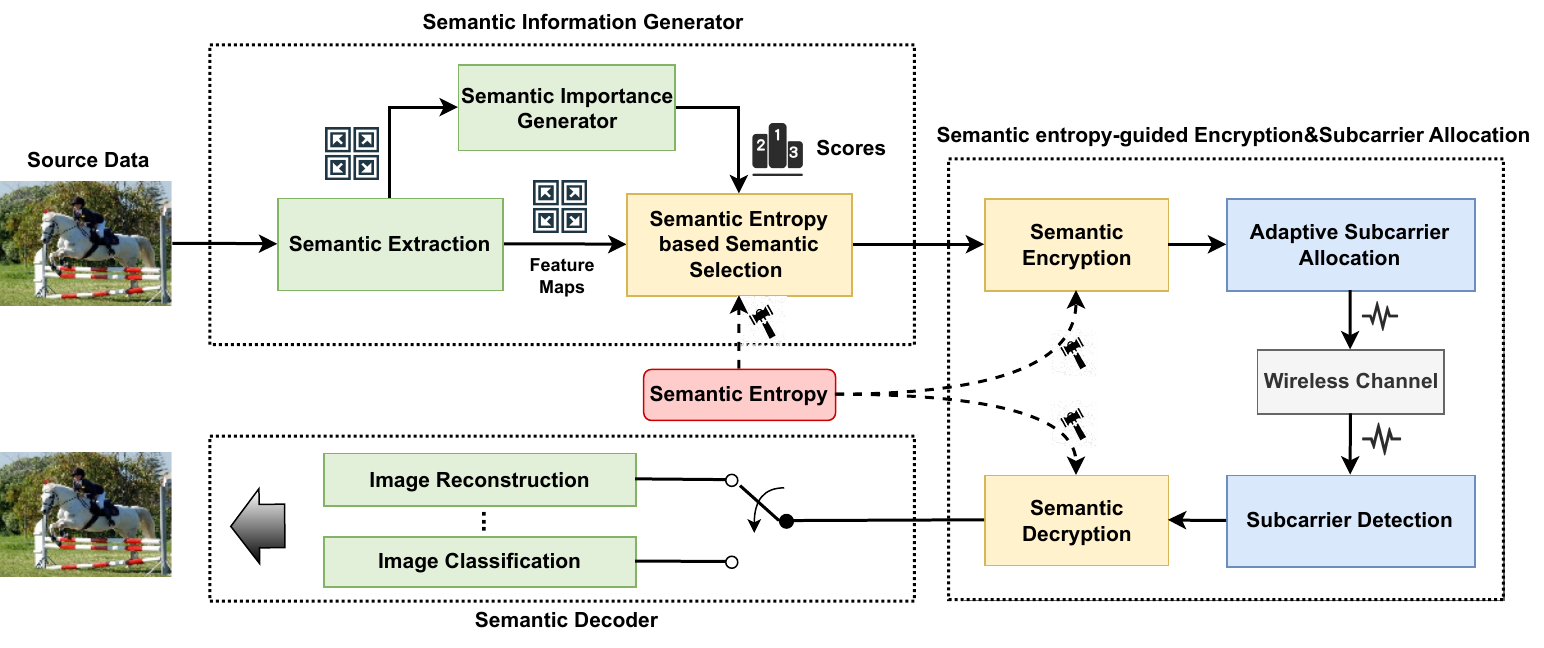}}
\caption{The general architecture of our SemEntropy. Controlling semantic selection and guiding semantic encryption and adaptive subcarrier allocation through semantic entropy.
}
\label{fig1}
\end{figure*}

\section{SemEntropy Architecture}
\subsection{System Model}
This paper introduces a novel semantic communication architecture for image transmission: SemEntropy. The system framework of SemEntropy is depicted in Fig. \ref{fig1}. Digital devices with computational capabilities capture the image or the sender inputs the image intended for transmission. Subsequently, the semantic extraction module extracts the semantic information through a carefully designed neural network. This process can be described by the function:
\begin{equation}
\mathbf{Feat}=En(\mathbf{p},\zeta)
\end{equation}
where $\mathbf{p}$ represents the source image, and $En(\cdot)$ is the semantic extraction network with trainable parameters $\zeta$. The output $\mathbf{Feat}$ has $N$ channels, each corresponding to a feature map. For specific communication tasks, the importance of feature maps corresponding to different channels varies. The extracted feature maps are processed through the semantic importance generator module, and their contribution to the task is calculated separately to generate their semantic scores. For the task $Y$, this process can be expressed as:
\begin{equation}
w_i=W(\mathbf{Feat},Y)
\end{equation}
where $W(\cdot)$ represents the semantic importance generator module and $w_i$ represents the semantic scores of the feature maps. We define semantic entropy for image semantic transmission based on the original semantic entropy, and designe a semantic selection module guided by semantic entropy. This process can be expressed as:
\begin{equation}
    \mathbf{F} = S(\mathbf{Feat} ,w_i,semantic\_entropy)
\end{equation}
where $S(\cdot)$ represents the semantic selection module, $\mathbf{F} = [F_1,F_2\ldots,F_\lambda
]$ are the selected $\lambda
$ feature maps.

Subsequently, Semantic entropy will guide physical layer encryption and adaptive subcarrier allocation.
In the OFDM system, the extracted semantic feature maps are encoded into OFDM signals with $x$ subcarriers, $[s_1,s_2\ldots,s_x]$. The adaptive subcarrier allocator $A(\cdot)$ assigns the OFDM signal to various subchannels based on CSI. This process can be expressed as:
\begin{equation}
s^{\prime}=A(s)
\end{equation}
where $s^{\prime}$ denotes the allocated sequence. Following the guidance of semantic entropy, we intricately encrypt information by combining semantic keys with physical layer keys to provide more precise protection. Once allocated and encrypted, these subcarriers are subjected to transmission over the wireless channel.
After the receiver captures the OFDM symbol, the cyclic prefix (CP) is removed, and the discrete Fourier transform (DFT) is applied to convert the time-domain signal back into the frequency domain. Afterwards, the receiver recovers the data based on the received semantic information and completes specific tasks.

\begin{figure*}[t]
\centerline{\includegraphics[width=1.0\linewidth]{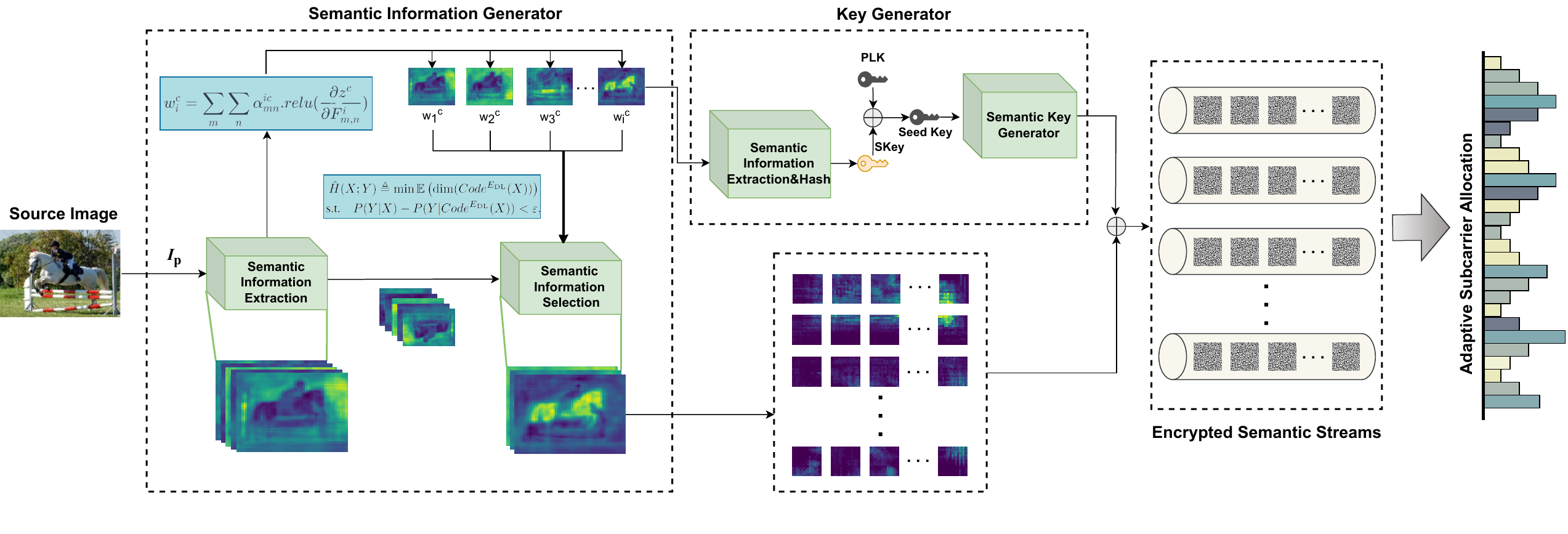}}
\caption{The detailed procedure of semantic information generator, semantic entropy-guided encryption and adaptive subcarrier allocation.}
\label{detail}
\end{figure*}

\subsection{Semantic Information Generator}
In our SemEntropy system, the semantic information generator includes three modules: semantic extraction, semantic importance generator and semantic selection.

As shown in Fig. \ref{detail}, the goal of semantic extraction is to distill high-dimensional features from pixel data and to conduct fine-grained analysis via annotations of semantic feature maps. In our proposed model, the input image is denoted as $I_p\in\mathbb{R}^{H\times W\times3}$, where $H$ and $W$ represent the height and width of the image, respectively. The image initially undergoes a feature mapping process, projecting it into the corresponding high-dimensional semantic space.
Through multi-layered feature mapping computations, hierarchical and rich feature information is progressively extracted from the image. After this stage, in order to further fine-grained annotation, the system integrates a flexible enhancement module that can be adjusted for specific tasks. This module uses adjustable dilation convolution to systematically expand the receptive field, enabling the capture of a wide range of spatial information without losing resolution. Through pooling strategy, features at different scales were further integrated. This ensures that our system can adapt to various challenges and maintain accuracy and detail in different task requirements. After processing through this network, we obtain a highly feature-rich feature map \(\mathbf{Feat} \in \mathbb{R}^{H' \times W' \times N}\). Here, \(H'\) and \(W'\) represent the spatial dimensions of the feature map, while \(N\) denotes the number of feature channels. The specifics of the model implementation can be found in Section \ref{sec:experiment}.

Another key function of the semantic information generator is semantic importance generator, which is highly task related. Tagged feature maps are divided into different semantic flows, each containing different semantic information. This complex relationship can be naturally represented by model weights in neural networks. If we consider the final output of the feature extraction module as \( z^c \), the significance of the $i$-th feature map concerning task $c$ can be symbolized as \( w^c_i \). The computation of \( w^c_i \) seamlessly integrates global average pooling with gradient backpropagation:

\begin{equation}
w_i^c=\sum_m\sum_n\alpha_{mn}^{ic}.relu(\frac{\partial z^c}{\partial F^i_{m,n}})
\end{equation}
where $\alpha_{mn}^{ic}$ represents the  gradient weights for a particular class $c$ and feature map $i$. \( F^i_{m,n} \) represents the activation value of the feature map located at the $m$-th row and $n$-th column. To prevent negative effects when $w^c$ is negative, we uniformly take the absolute value $|w_i^{c}|$ of $w_i^c$.

\subsection{Semantic Entropy}
For different tasks, the contribution of the feature maps is not consistent, and it's unnecessary to transmit all feature maps. The precise selection of features becomes particularly crucial. To accurately select feature maps, we introduce semantic entropy in the image semantic transmission tasks. Compared to the information entropy based on statistical features of the source symbols, semantic entropy directly quantifies the semantic information of the source. Although a general concept of semantic entropy has been proposed \cite{entropy1,entropy2}, there is still lack a practicable quantification method in the context of image semantic transmission. According to the general definition of entropy, semantic entropy is a measure concerning the source and task:

\noindent \textbf{Definition 1.} \textit{Given semantic source $\mathcal{X}$, \textit{semantic entropy} is defined as the minimum expected number of semantic symbols about the data $X \in \mathcal{X}$ that are sufficient to predict task $Y$, i.e.,}
\begin{equation}
\begin{gathered}
H(X;Y) \triangleq \min_{E_S} \mathbb{E}\left(\text{dim}(Code^{E_S}(X))\right) \\
\text{s.t.}\quad P(Y|Code^{E_S}(X)) = P(Y|X),
\end{gathered}
\end{equation}
\textit{where $Code^{E_S}(X)$ denotes the semantic symbol vector extracted from $X$ with the semantic encoder $E_S$, and $P(Y|X)$ is the conditional probability of $Y$ given $X$.}

From Definition 1, the semantic entropy of $X$ given $Y$ is essentially defined as the expected value over the entire dataset $\mathcal{X}$. Hence, for the same task and dataset, the semantic entropy remains a constant. In pursuit of a meaningful and actionable measure, \cite{entropy1} employs a well-designed deep learning (DL) model as a semantic encoder to approximate the semantic entropy for the task, denoted as:

\begin{equation}
\begin{gathered}
\tilde{H}(X;Y) \triangleq\min\mathbb{E}\left(\dim(Code^{E_{\mathrm{DL}}}(X))\right)\\
\text{s.t.}\quad P(Y|X)-P(Y|Code^{E_{\mathrm{DL}}}(X))<\varepsilon,
\end{gathered}
\end{equation}

where the constraint indicates that the gap between $P(Y|X)$ and $P(Y|Code^{E_{\mathrm{DL}}}(X))$ can not exceed $\varepsilon$.

The above formula definitions and formulas provide a qualitative analysis of semantic entropy. However, in practical operation and use, there has not been sufficient research on how to quantitatively analyze and calculate semantic entropy. In the context of a semantic communication system for image semantic transmission, we represent the dataset $X$ with a set of feature maps $\mathbf{Feat}$ obtained after feature extraction module. Within this framework, the semantic entropy $H(X;Y)$ is essentially defined as the minimal expected number of feature maps in $F$ that are necessary to accomplish task $Y$. Specifically, this notion can be articulated and calculated in the following manner:
\begin{equation}
\begin{gathered}
\tilde{H}(X;Y) \triangleq\min\mathbb{E}\left(cnt(\mathbf{Feat})\right)\\
\text{s.t.}\quad P(Y|X)-\sum_{i=0}^{\lambda} w_i^c<\varepsilon,
\end{gathered}
\end{equation}

This  formula endows semantic entropy with practical and operational value, making it an effective measure for filtering and transmitting crucial feature mappings in real-world tasks. In this way, we can precisely address challenges in the real world, enhancing the efficiency and accuracy of task execution.

\subsection{Semantic Entropy-Guided Encryption and Adaptive Subcarrier Allocation}
SemEntropy sufficiently explores the semantic properties of data. Given the uncertainty of the source and the dynamic nature of task measurement by neural networks, semantic entropy inherently exhibits ample randomness. It does not rely on additional relays to generate randomness, giving semantic entropy a significant advantage in terms of complexity and security. Furthermore, through the integrated design of semantic entropy and adaptive subcarrier allocation, we not only enhance the efficiency of data transmission but also fortify the system's robustness across diverse semantic communication scenarios. This collaborative optimization strategy enables the system to intelligently adjust subcarrier allocation in dynamic environments, adapting to the varying demands of different tasks for communication quality and security.

\begin{figure*}[t]
\centerline{\includegraphics[width=1.0\linewidth]{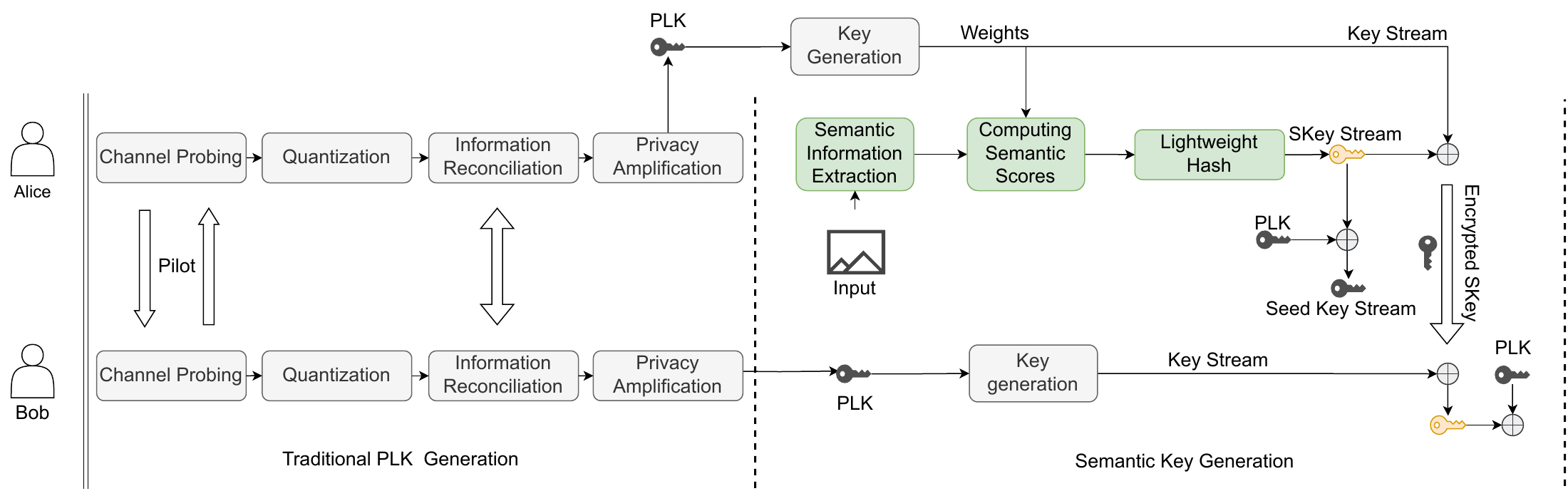}}
\caption{The process of generating and exchanging semantic keys. We take the semantic scores generated by the semantic information generator and the semantic key SKey generated by traditional PLK.  }
\label{encryption}
\vspace{-1mm}
\end{figure*}

Semantic entropy-guided encryption aims to compensate for the vulnerability of wireless channels in static fading environments, ensuring effective protection of data confidentiality and integrity even at low key generation rates. As shown in Fig. \ref{encryption}, as an image is input into the semantic information generator deployed on the transmitter, the semantic scores of various feature maps $w = [w_1,w_2,\ldots,w_i]$ can be extracted under the constraint of semantic entropy. 
Using the key stream $k_{stream}$, we then compute the weights of each semantic score. Finally, we generate the semantic keys by inputting the weighted sum of $W$. Such a procedure can be expressed as:

\begin{equation}
SKey_{stream}=HASH(w_i \times W(p))
\end{equation}
where $W(\cdot)$ represents the calculation of importance scores, and $HASH$ refers to lightweight hash algorithm.  Once $SKey_{stream}$ is obtained, it is encrypted using $PLK$ and transmitted to the other side of the communication channel. The receiving end only needs to employ the $PLK$ to decrypt the encrypted $SKey_{stream}$ to retrieve $SKey_{stream}$, thereby sharing it between the two endpoints. 

On the sender's side, the $SKey_{stream}$ is utilized to semantically encrypt the transmitted information.  Subsequently, by the operation $SKey_{stream} \oplus PLK$, we derive the seed key stream and further produce an encryption/decryption key stream. To ensure security, the $SKey_{stream}$ is encrypted using the conventional $PLK$ stream before being transmitted to the authorized recipient. This encrypted keys are then used for decryption operations at the receiving end.

According to the reciprocity of the channel, the receiving end can obtain the noise power of the subchannel. For example, frequency selection channels can be divided into subchannels with different Signal-to-noise ratios (SNR). The receiving end estimates the CSI of all subchannels and shares it with the sending end. SemEntropy utilizes these channel conditions at the technical level, which helps to protect the most important key points while improving transmission accuracy. The receiving end sorts the subchannels based on channel conditions and feeds back this sequence to the sending end. 
In OFDM systems, modulated signals are represented as $s_n$. Acknowledging the existence of multi-path channels in OFDM, each modulated bit $s_n$ is dispatched to a designated subchannel  $h_n$, preparing it for transmission. This modular structure facilitates the concurrent transmission of several modulated bits across different subchannels. Representing the channel gains in a wireless multi-path scenario, we have:
\begin{equation}
\mathbf{H}_c = [h_1, h_2, \ldots , h_{N_c}]^T 
\end{equation}

Here, $N_c$  defines the overall count of subchannels within ${H}_c$, and each $h_n$ delineates the gain linked to the $n^{th}$ subchannel.
Transferring information with high importance scores on better subchannels can achieve more reliable transmission of semantic information. To this end, an ascending sorting method is used to establish a mapping function $\mathcal{M}$ between semantic information and each sub channel. This mapping relies on the weights calculated for semantic information and CSI. The higher the weight is, the more reliable the subchannels to be assigned are. The mapping function is represented as:
\begin{equation}
\mathcal{M}(w_i,h_c)=\{{I}_{i},h_j\},
\end{equation}
where the map $ \{{I}_{i},h_j\} $ refers to transmit the semantic information $I_i$ at the subchannel $h_j$. Before transmission, diverse semantic information will be encrypted using distinct SKeys. 

Each OFDM symbol is processed through the Inverse Discrete Fourier Transform (IDFT) and combined with a CP to create the transmission signal $\mathbf{y}$. Upon reception, the noisy signal, $\hat{\mathbf{y}}$, is cleansed of the CP and subjected to the DFT to extract the frequency domain pilots, $\hat{\mathbf{Y}_p}$, and data symbols, $\hat{\mathbf{Y}}$. These operations uphold the equations presented for both the pilot and information symbols,
\begin{equation}
\label{eq16}
\begin{aligned}\hat{\mathbf{Y}_{p}}[i,k]&=H[k]{\mathbf{Y}_{p}}[i,k]+W[i,k],\\\hat{\mathbf{Y}}[j,k]&=H[k]{\mathbf{Y}}[j,k]+V[j,k],\end{aligned}
\end{equation}
where $H[k]$ illustrates the channel's frequency response for the $\text{k-th}$ subcarrier, and $W$ as well as $V$ denote the noise samples. Then, the decoder will complete the subsequent operations with given $\mathbf{Y}_{p}$, $\hat{\mathbf{Y}}_{p}$ and $\hat{\mathbf{Y}}$. Algorithm 1 describes the procedure of semantic entropy-guided encryption and adaptive subcarrier allocation.

\begin{algorithm}[t]
\caption{Semantic Entropy-Guided Encryption and Adaptive Subcarrier Allocation}
\hspace*{0.02in} {\bf INPUT:} 
image p, seed(PLK or (SKey  $\oplus$ PLK  )), semantic entropy, $N_c$
\begin{algorithmic}[1]
\State  $F$ = $[F_1,F_2,\dots,F_N]$ = Semantic\_Generation(p, semantic entropy) 
\State  $w_i$ = Importance\_Modeling(F, semantic entropy)
\State  $weight_i$ = Weight\_generation(seed)
\State  Generated\_scores = $weight_{i} \times w_{i} \pmod{1}$ 
\State  $SKey_{stream}$ = Lightweight\_Hash(Generated\_$scores_i$) 
\State  CSI estimation: $H_c$ = $[h_1, h_2, \ldots , h_{N_c}]^T  $
\For{ i $\leftarrow$ 1 to N}
    \State  $F_i^{code}$ = channel\_code$(F_i)$
    \State  $F_i^{enc}$ = semantic\_encrypt($F_i, SKey_i$)
    \State  $SubLoc_i$ = Subcarrier\_Allocation($F_i^{enc}$, $w_i$, $H_c$)
\EndFor
\end{algorithmic}
\label{alg:key}
\end{algorithm}


\section{Experiments and Numerical Results} \label{sec:experiment}
This section will provide a detailed introduction to the experimental setup, experimental dataset, and experimental results. The transmission performance of SemEntropy was evaluated through experiments on two application scenarios: image reconstruction and image classification, and the security of the encryption scheme was analyzed.

\begin{table}[htbp]
\centering
\renewcommand{\arraystretch}{0.9}
\caption{The semantic information generator architecture for classification tasks.}
\begin{tabular}{|c|c|c|c|}
\hline
Layer Name & Output Size & Details &Number \\
\hline
Initial Block & \(128 \times 128\) & \(7 \times 7, 64, \text{ stride } 2\) & $\times 1$\\
& & \(3 \times 3\), max pool, stride 2 & \\
\hline
Layer 1 & $128 \times 128$ & $1 \times 1, 64, stride 1$ &\\
& & $3 \times 3$, $64, stride 1$ & $\times 3$\\
& & $1 \times 1$, $256, stride 1$ & \\

\hline
Layer 2 & $64 \times 64$ & $1 \times 1, 128, stride 2$ &\\
& & $3 \times 3$, $128, stride 1$ & $\times 4$\\
& & $1 \times 1$, $512, stride 1$ & \\

\hline
Layer 3 & $64 \times 64$ & $1 \times 1, 256, stride 1$ &\\
& & $3 \times 3$, $256, stride 1$ & $\times 6$\\
& & $1 \times 1$, $1024, stride 1$ & \\
\hline
Layer 4 & $64 \times 64$ & $1 \times 1, 512, stride 1$ &\\
& & $3 \times 3$, $5126, stride 1$ & $\times 3$\\
& & $1 \times 1$, $2048, stride 1$ & \\
\hline
ASPP & \(64 \times 64\) & $3 \times 3$, 256,  stride  1  & \\
& & AdaptiveAvgPool2d & \\
& & $1 \times 1$, 1280, \text{ stride } 1 & $\times 3$ \\
& & Interpolate &  \\
& & $1 \times 1$, 256, \text{ stride } 1 &  \\
& & Dropout(0.5) & \\
\hline
Final conv & $64 \times 64$ & $1 \times 1$, 256, stride 1 & $\times 1$\\
& & $1 \times 1$, 21, stride 1 & \\
\hline
\end{tabular}
\label{sct}
\vspace{-2mm}
\end{table}

\begin{table}[htbp]
\centering
\renewcommand{\arraystretch}{0.9}
\caption{The semantic information generator architecture for reconstruction tasks.}
\begin{tabular}{|c|c|c|c|}
\hline
Layer Name & Output Size & Details &Number \\
\hline
Conv2d & $32 \times 32$ & $ 5 \times 5$, 64,  stride 2 & $\times 1$ \\  
\hline
Conv2d & $16 \times 16$ & $ 3 \times 3$, 128,  stride 2 & $\times 1$ \\ (downsampling) & & &  \\
\hline
Conv2d & $8 \times 8$ & $ 3 \times 3$, 256,  stride 2 & $\times 1$ \\
(downsampling) & & & \\ 
\hline
Residual Block & $8 \times 8$ & $ 3 \times 3$, 256,  stride 2 & $\times 1$ \\
& & $ 3 \times 3$, 256,  stride 2 &  \\
\hline
Conv2d & $8 \times 8$ & $ 3 \times 3$, N,  stride 1 & $\times 1$ \\

\hline
\end{tabular}
\label{sct2}
\end{table}

\subsection{Model Structure}
\subsubsection{Semantic Information Generator}
The semantic information generator is used to extract semantic information and calculate importance scores, where the semantic extraction module for classification tasks and reconstruction tasks are shown in Table.\ref{sct} and Table.\ref{sct2}, respectively. 

For classification tasks, we
take the output of the last convolutional layer as the feature maps, also known as the semantic label map, to be selected, as the number of feature maps here is the same as the number of target classes and each of these maps and classes corresponds to each other one by one, which directly contribute to the final result. 

The semantic extraction network for reconstruction tasks is shown in Table.\ref{sct2}. It consists of a normalization layer and a series of down-scaling convolutional layers known as the down-sampling layers and residual blocks. The output (dimension is $N\times\frac{H}{2^{d}}\times\frac{W}{2^{d}}$, $\text{d}$ is the down-sampling factor) of the feature extraction for input image(dimension of $C\times H\times W$) also consists of $N$ channels.

\subsubsection{Semantic Decoder}
\begin{figure}[t]
\centerline{\includegraphics[width=0.7\linewidth]{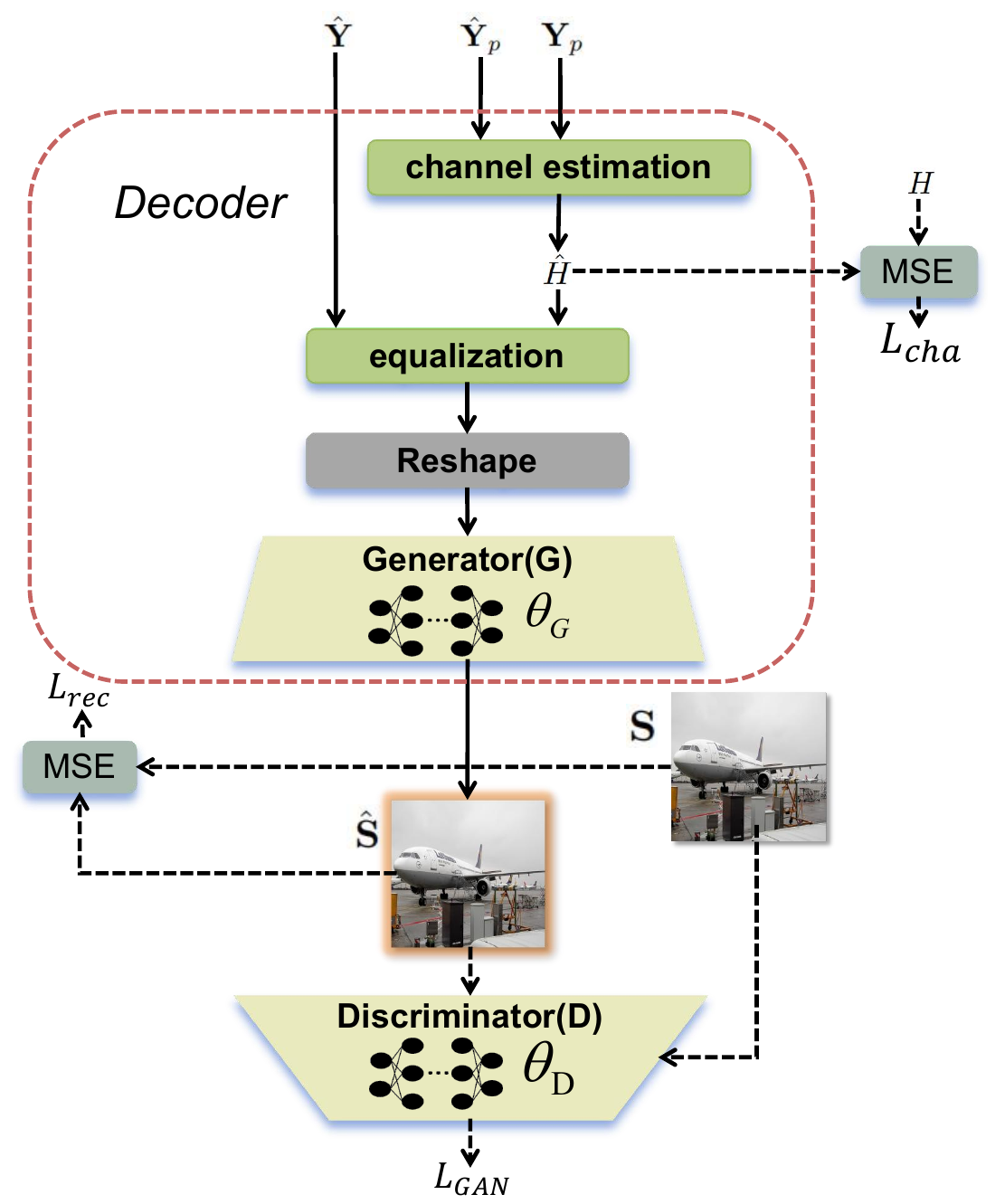}}
\caption{The inner architecture of the decoder. }
\label{decoder}
\vspace{-3mm}
\end{figure}

For reconstruction tasks the decoder is trained to estimate transmitted source $\hat{\mathbf{S}}=$$De^{rec}(\mathbf{Y}_{p},\hat{\mathbf{Y}},\hat{\mathbf{Y}}_{p}:\theta_{rec})$, given $\mathbf{Y}_{p}$, $\hat{\mathbf{Y}}_{p}$ and $\hat{\mathbf{Y}}$, where $De^{rec}(\cdot)$ is the semantic decoder for reconstruction tasks and $\theta_{rec}$ is the trainable parameters. It is composed up of four parts, including the channel estimation, the equalization, the generator and the discriminator. Note that only the generator and discriminator need to be trained. To estimate the source information we  adopt the per-channel minimum mean square error (MMSE) channel estimation method, which is as follows:
\begin{equation}
\label{eq18}
H_{MMSE}\left[k\right]=\frac{\sum_{i=1}^{N_{p}}\hat{\mathbf{Y}}_{p}[i,k]\mathbf{Y}_{p}[i,k]^{*}}{N_{p}+\sigma^{2}}.
\end{equation}
The pilot symbols $\mathbf{Y}_{p}[i,k]$ are designed to have unit power. 
The channel estimation and equalization are included as pre-processing steps in front of the generator network $G$ in the decoder as shown in Fig. \ref{decoder}.
\begin{equation}
\label{eq19}
\mathbf{Y}_{MMSE}[k]=\frac{\mathbf{Y}[i,k]\hat{H}[i,k]^{*}}{|\hat{H}[i,k]|^{2}+\sigma^{2}}.
\end{equation}
The structure of the generator network is almost symmetric (in the reverse order) to the feature extraction network for reconstruction, except that the final activation function is set to Sigmoid function to enforce a valid dynamic range of image output pixels. Here, it can be expressed as:
\begin{equation}
\hat{\mathbf{S}}=G(\mathbf{Y}_{MMSE}:\theta_{G}\subseteq\theta_{rec})
\end{equation}
where $G(\cdot)$ is the generator and $\theta_G$ is the trainable parameters.

We extend our model to an adversarial setting where an additional discriminator neural network is introduced to distinguish whether the image is original or generated by the decoder. The discriminator can be expressed as:
\begin{equation}
Dis(\mathbf{\cdot})=D(\mathbf{S},\hat{\mathbf{S}}:\theta_{D}\subseteq\theta_{rec}) 
 \\Dis(\cdot)\in[0,1]
\end{equation}
where $Dis(\cdot)$ is the output of the discriminator for input image, $D(\cdot)$ is the decoder’s discriminator and $\theta_{D}$ is the trainable parameters.

For the semantic classification tasks, our decoder’s goal is to generate the mask corresponding to the original image based on the results, $\mathbf{Y}_{p}$, $\hat{\mathbf{Y}}_{p}$ and $\hat{\mathbf{Y}}$, obtained from the OFDM receiver, which can be expressed as: 
\begin{equation}
Mask=De^{seg}(\mathbf{Y}_{p},\hat{\mathbf{Y}},\hat{\mathbf{Y}}_{p}:\theta_{seg})
\end{equation}
where $De^{seg}(\cdot)$ is the semantic decoder for classification tasks and $\theta_{seg}$ is the trainable parameters.
Since in the feature extraction module we use the output of the last convolutional layer as the object of semantic selected, and this output needs to be up-sampled once so that the output size of the mask is the same as the original image input size. So in the decoder we have an bilinear-interpolation-based upsampling layer, which uses the four nearest pixel points to estimate the pixel value at the new location, and a softmax layer to get the final mask.

\subsubsection{Losses}
The loss functions consists of a softmax cross entropy loss used for classification tasks, a reconstruction loss and a gan loss used for reconstruction tasks, and a channel estimation loss for both tasks. 

For channel estimation and equalization module, we introduce an additional channel estimation loss term to guarantee that the input for equalization is a valid channel estimation. The reconstruction loss can then be expressed as follows:

\begin{equation}
    \begin{aligned}L_{cha}& =\mathbb{E}\big[||\hat{H}-H||_{2}^{2}\big]  \\&=\mathbb{E}\Big[||H_{MMSE}-H||_{2}^{2}\Big].\end{aligned}
\end{equation}

The reconstruction loss for reconstruction tasks can be expressed as follows:
\begin{equation}
L_{rec}=\mathbb{E}_{p(\mathbf{S},\hat{\mathbf{Y}},\hat{\mathbf{Y}}_{p})}\big[||De^{rec}(\mathbf{Y}_{p},\hat{\mathbf{Y}},\hat{\mathbf{Y}}_{p})-\mathbf{S}||_{2}^{2}\big],
\end{equation}
where $\mathbb{E}_{p}$[] is the expected value over distribution $p$ and $De(\mathbf{Y}_{p},\hat{\mathbf{Y}},\hat{\mathbf{Y}}_{p})(=\hat{\mathbf{S}})$. Here, we use mean squared error (MSE) loss as our distortion function. The adversarial loss is defined in eq. \ref{eq25} where $Dis(\mathbf{S})$ is the discriminator network output for input image $\mathbf{S}$ (desired output is 1 for uncompressed images and 0 for images generated by the decoder).
\begin{equation}
\label{eq25}
L_{GAN}=\max_{Dis}\mathbb{E}[(Dis(\hat{\mathbf{S}})-1)^2]+\mathbb{E}[Dis(\mathbf{S})^2].
\end{equation}
For the proposed end-to-end system, we jointly optimize the
reconstruction loss, channel estimation loss and the adversarial loss. When the adversarial loss is adopted, the training process becomes a minmax optimization and the objective function is modified as follows where $\lambda_g$ denotes the weight for the adversarial loss and $\lambda_c$ represents the weight for channel estimation loss. 
\begin{equation}
\label{eq26}
\min\limits_{E,D}L_{total}=L_{rec}+\lambda_{c}L_{cha}+\lambda_{g}L_{GAN}.
\end{equation}

\begin{figure}[t]
\centerline{\includegraphics[width=0.7\linewidth]{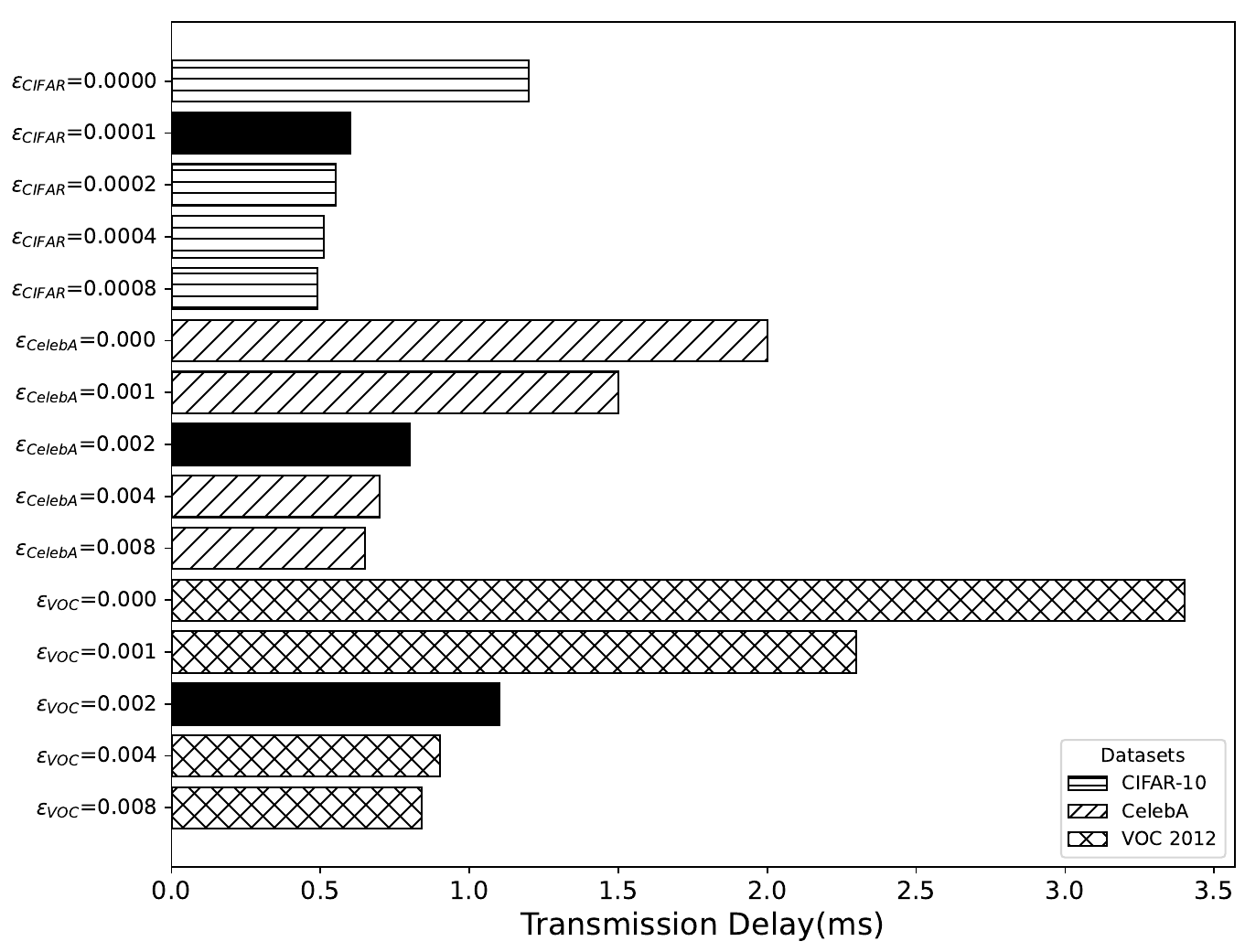}}
\caption{Effect of $\epsilon$ change on transmission delay for reconstruction tasks. }
\label{epsilon_rec}
\end{figure}

\begin{figure}[t]
\centerline{\includegraphics[width=0.7\linewidth]{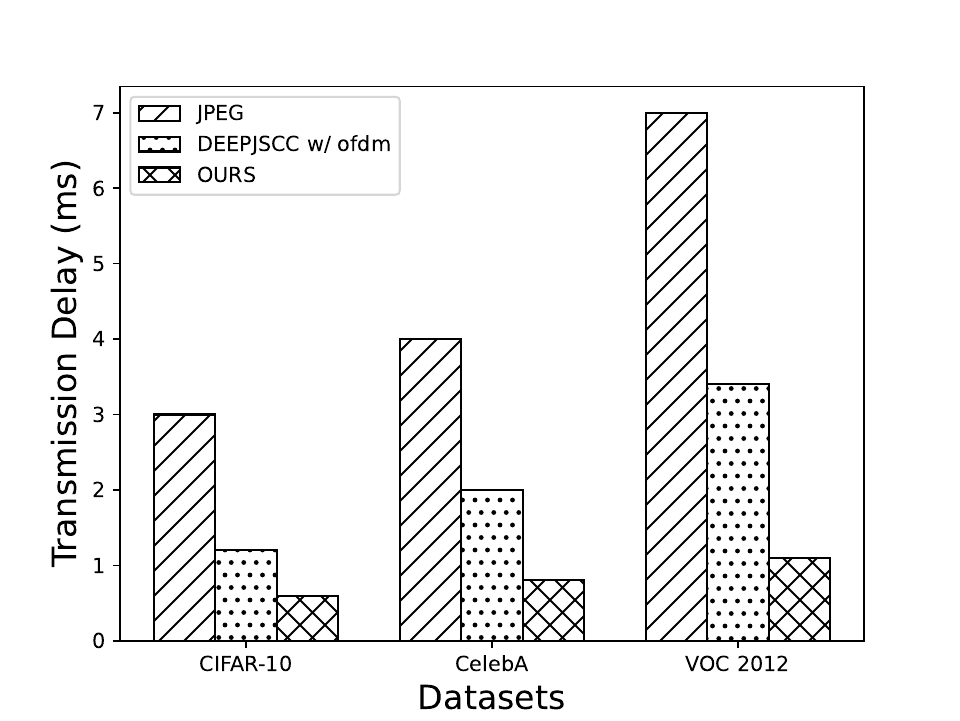}}
\caption{Transmission Latency for reconstruction tasks. }
\label{delay_rec}
\end{figure}

\begin{figure*}[t]
\vspace{-5mm}
\centering

\subfloat[]{
\includegraphics[width=2in]{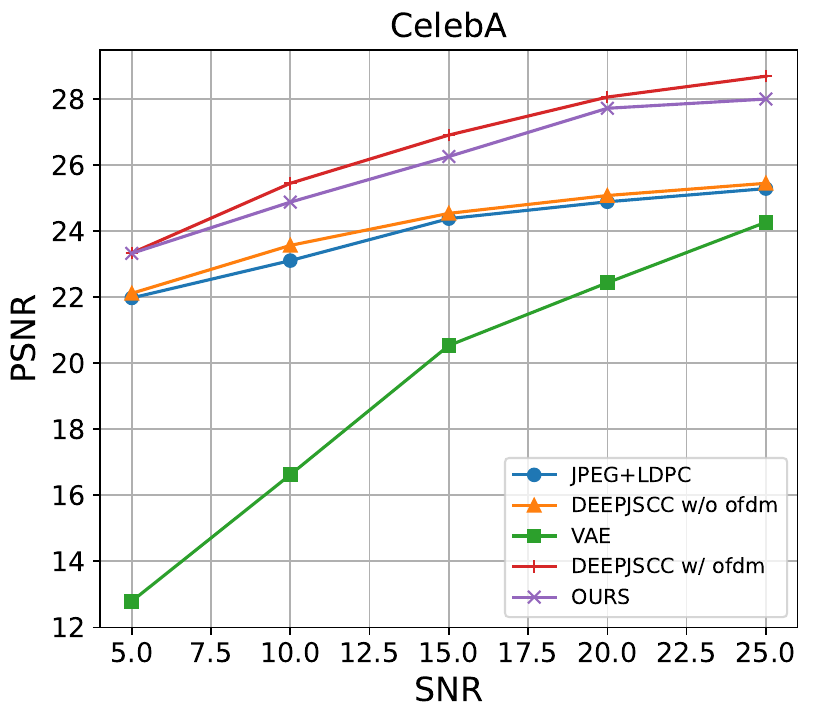}
\label{fig:CelebA-psnr}
}
\subfloat[]{
\includegraphics[width=2in]{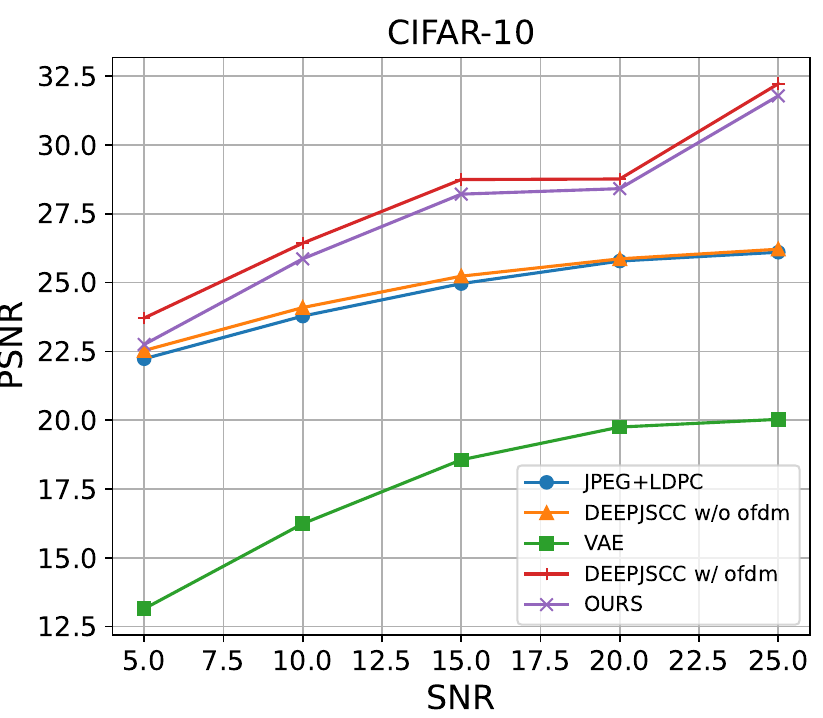}
\label{fig:CIFAR10-psnr}
}
\subfloat[]{
\includegraphics[width=2in]{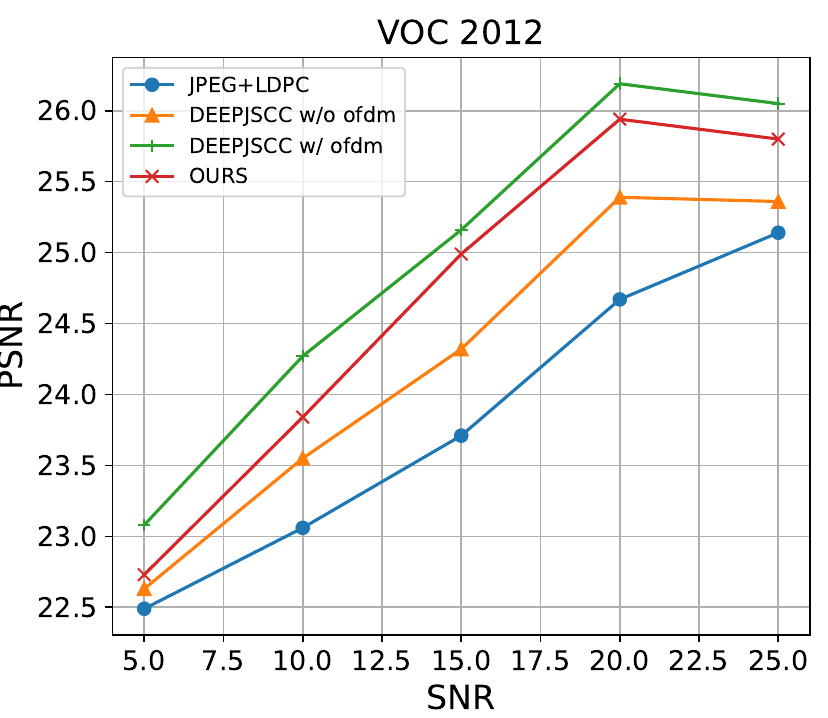}
\label{fig:VOC-psnr}
}

\vspace{-3mm}

\subfloat[]{
\includegraphics[width=2in]{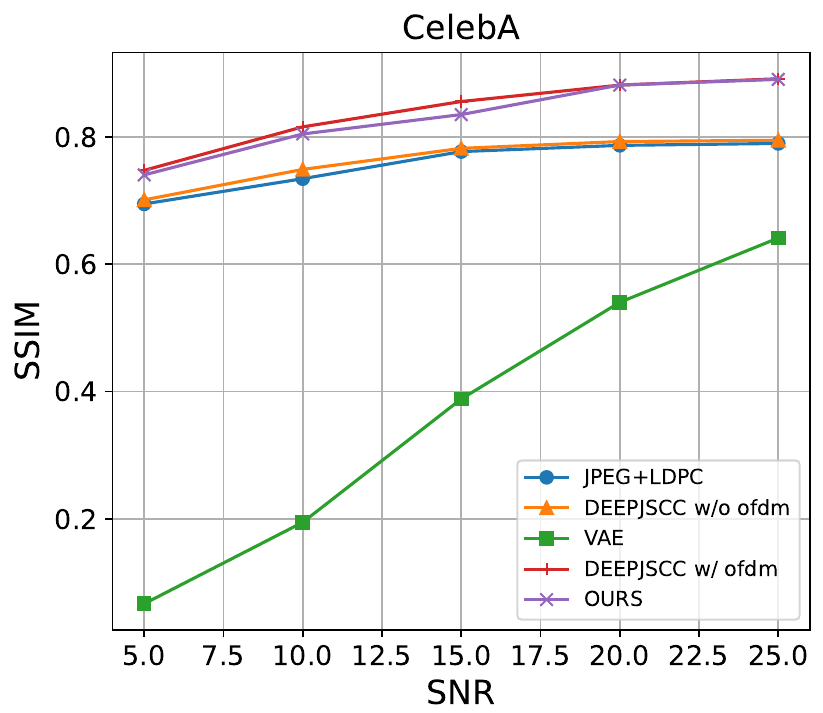}
\label{fig:CelebA-ssim}
}
\subfloat[]{
\includegraphics[width=2in]{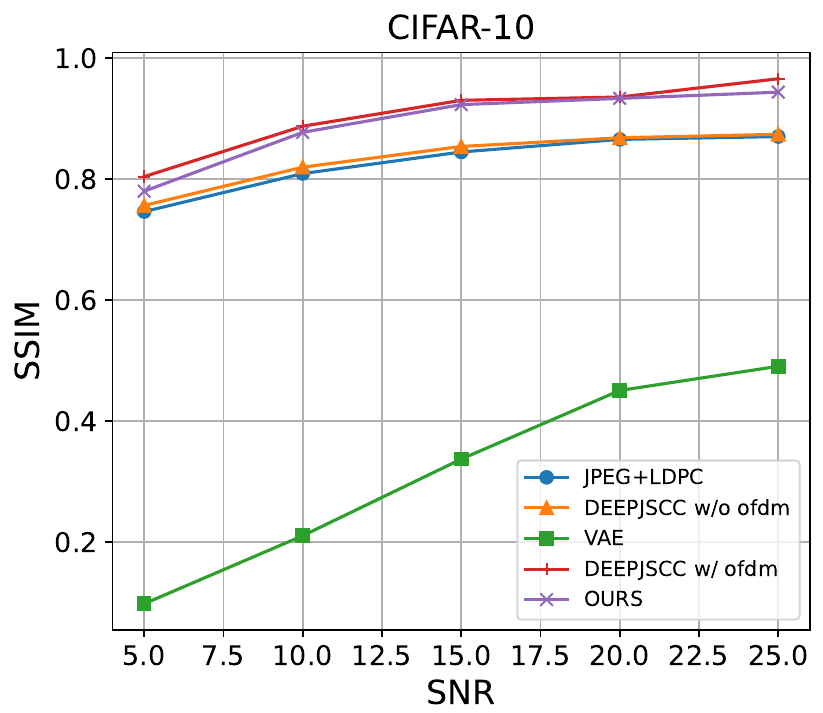}
\label{fig:CIFAR10-ssim}
}
\subfloat[]{
\includegraphics[width=2in]{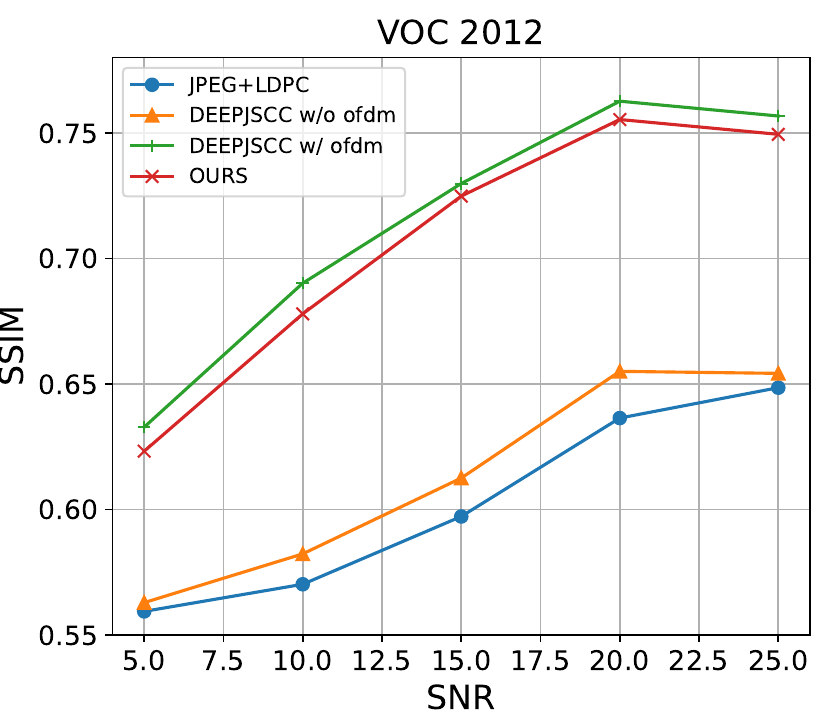}
\label{fig:VOC-ssim}
}

\caption{Performance comparison  under different SNRs and datasets for reconstruction tasks.}
\label{fig:psnr-ssim}
\vspace{-2mm}
\end{figure*}

For the classification tasks, we jointly optimize the
softmax cross entropy loss and channel estimation loss. The total loss is modified from eq. \ref{eq26} to eq. \ref{eq23}:
\begin{equation}
\label{eq23}
\begin{aligned}
L_{total}&=L_{Softmax}+\lambda_{c}L_{cha}
\\& = -\frac{1}{N}\sum_{i=1}^{N}log(\frac{e^{z_i}}{\sum_{j=1}^{|C|}e^{z_j}}) + \lambda_{c}L_{cha}
\end{aligned}
\end{equation}
where $N$ is the batch size, $C$ is the number of classes, and $z_{i}$ represents the value of each pixel from the generated segmentation mask.

\subsection{Settings and Datasets}
We optimize the loss function in eq.(\ref{eq23}) and set $\lambda_c = 0.5$ for all experiments and $\lambda_g = 2\times10^{-4}$ for 5dB SNR and  $\lambda_g = 5\times10^{-5}$ for 15dB SNR for reconstruction experiments. We use a batch size of 128 for CIFAR10, a batch size of 64 for CelebA, and a batch size of 16 for VOC2012. All neural networks are trained using ADAM with $\beta_{1}=0.5,\beta_{2}=0.999$, and an initial learning rate of $5\times10^{-4}$. We train for 300 epochs, 30 epochs, and 40 epochs for CIFAR10, CelebA and VOC2012 respectively. For all datasets, we apply linear learning rate decaying for the second half of the training process.  Specifically, the computer used for the experiment is equipped with a 13th Gen Intel(R) Core(TM) i9-13900K, 32G*4 RAM, and an NVIDIA GeForce RTX 4090.

\begin{figure}[t]
\centerline{\includegraphics[width=1.0\linewidth]{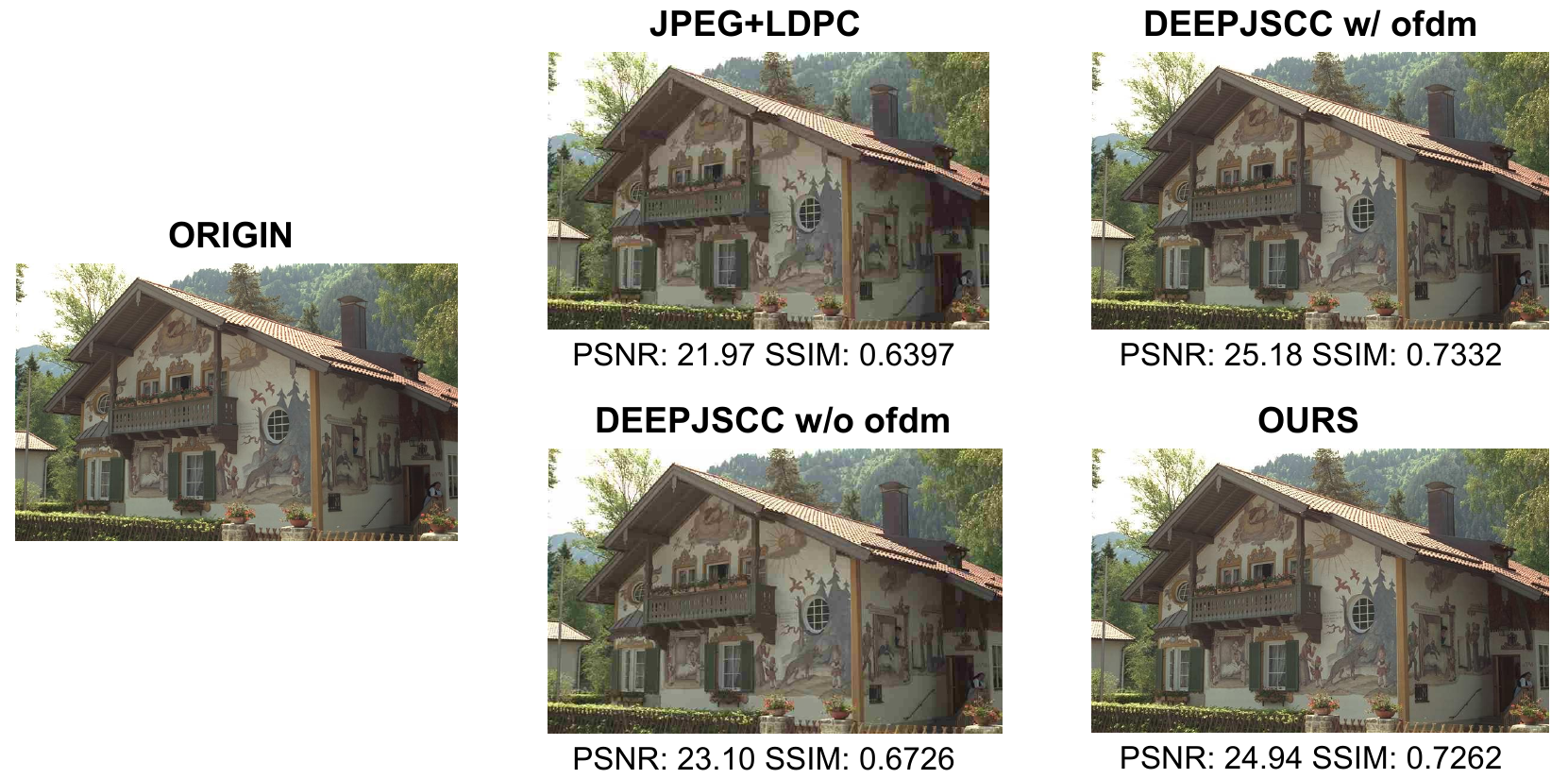}}
\caption{Comparison of reconstructed images. }
\label{kodak}
\vspace{-3mm}
\end{figure}

For reconstruction tasks, we first test our proposed method with images in CIFAR10, CelebA datasets for wireless communication. The CIFAR10 dataset contains 60,000 $32\times32$-pixel RGB images. Among the dataset, 50,000 images are used for training and the rest 10,000 images are used for testing. CelebA contains more than 200,000 celebrity images and these images are tagged with more than 40 features. Among the dataset, 162,770 images are used for training and 19,867 images are used for testing. Despite the small image size described here, our scheme can also be applied to high-resolution images, such as VOC2012. The PASCAL-VOC2012 dataset consists of 2,913 RGB images with sizes $512\times 512$ in 20 categories. Among the dataset, 10,582 images are used for training and the rest 1,449 images are used for testing. As the pre-processing, we adopt reflection padding and random cropping for CIFAR10, and we scale and crop CelebA images to $64\times 64$ pixels. For VOC2012, we will first pad each image to $512\times 512$, with filled value 0 and then randomly crop them into $480\times 480$. Then we normalize the data within the range of $[-1,1]$. 

For segmentation tasks, we test our proposed method with images in VOC2012 datasets and COCO datasets for wireless communication. The COCO dataset is a large, rich object detection, segmentation and captioning dataset. It offers 80 categories with over 330,000 images, 200,000 of which are labeled. In our experiments, we train and evaluate a subset of COCO 2017, on the 20 categories that are present in the PASCAL-VOC2012 dataset.

\subsection{Experiments for Reconstruction Tasks}
\subsubsection{Evaluation Metrics}
We utilize several metrics to evaluate the performance of our approach, notably the Peak Signal to Noise Ratio (PSNR) and the Structural Similarity Index Measure (SSIM):

\subsubsection{Baselines}
For better performance comparison, we used traditional JPEG compression combined with LDPC encoding, deep learning based DeepJSCC, and VAE as benchmarks to compare with our method. The following are specific descriptions of these methods:
\begin{itemize}
    \item \textbf{JPEG+LDPC}: This method combines two widely used techniques: JPEG image compression and LDPC channel encoding. 
    \item \textbf{DeepJSCC with/without OFDM}: This is a technology specifically designed for wireless image transmission, which implements joint source and channel encoding based on autoencoders.
    \item \textbf{VAE}:  The variational auto-encoder (VAE)-based semantic communication [58]. Note that we choose VAE instead of a standard autoencoder since VAE can better respond to the effects of noise.
\end{itemize}

Our initial experiment focuse on evaluating the impact of semantic selection parameter $ \epsilon $ on transmission delay, where $ \epsilon = 0 $ indicates that semantic selection was not used. Fig. \ref {epsilon_rec} indicates that as $\epsilon$ continues to increase, the transmission delay significantly decreases. This significant decrease indicates that semantic selection can effectively reduce transmission time. Based on this experimental result, we will select the $\epsilon$ values of the black markers in the graph for our future experiments. Specifically, select $0.0001$ for the CIFAR10 dataset, and $0.002$ for the CelebA and VOC2012 datasets.

Fig. \ref{delay_rec} compares the complexity of SemEntropy and baseline schemes from the perspective of the system runtime. SemEntropy has a lower transmission delay compared to the baselines. Experimental results on the VOC2012 dataset show that SemEntropy's transmission time is only $14\%$ of that of the traditional JPEG+LDPC scheme, which is $30\%$ of that of the DeepJSCC scheme combined with OFDM. Even on the CIFAR10 dataset, SemEntropy still showed excellent time performance. This is because SemEntropy uses a feature extraction network to extract the semantic information of the image, and further compresses the feature map through weight assignment and semantic entropy selection. We observe that SemEntropy is particularly suitable for systems with sensitive transmission delay and limited computing power. And the larger the data volume, the better the transmission effect of SemEntropy.

We then test the transmission performance of our scheme and baseline schemes under different SNR ratios. For more comprehensive testing, the experiment simulated semantic wireless transmission on the CIFAR10, CelebA and VOC2012 datasets, using SemEntropy and baseline schemes at a bit rate of approximately 0.15 at 16QAM.

\begin{figure*}[t]
\centerline{\includegraphics[width=0.8\linewidth]{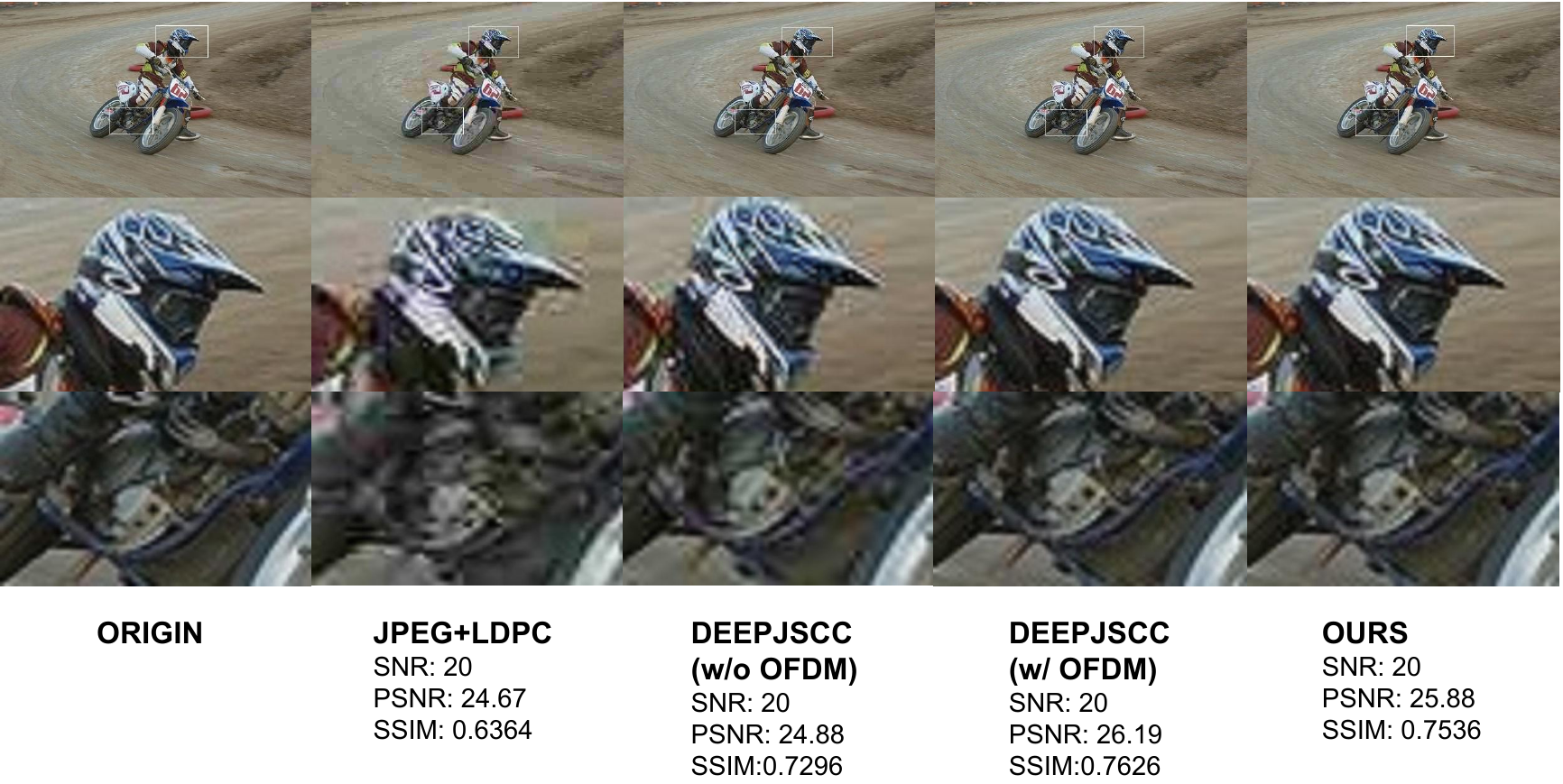}}
\vspace{-1mm}
\caption{Comparison of transmission effects of different schemes under the same transmission conditions. Note SemEntropy retains better detail information (helmet pattern color and motorcycle).}
\label{casestudy}
\end{figure*}

\begin{figure}[t]
    \centering  
    \includegraphics[width=0.7\linewidth]{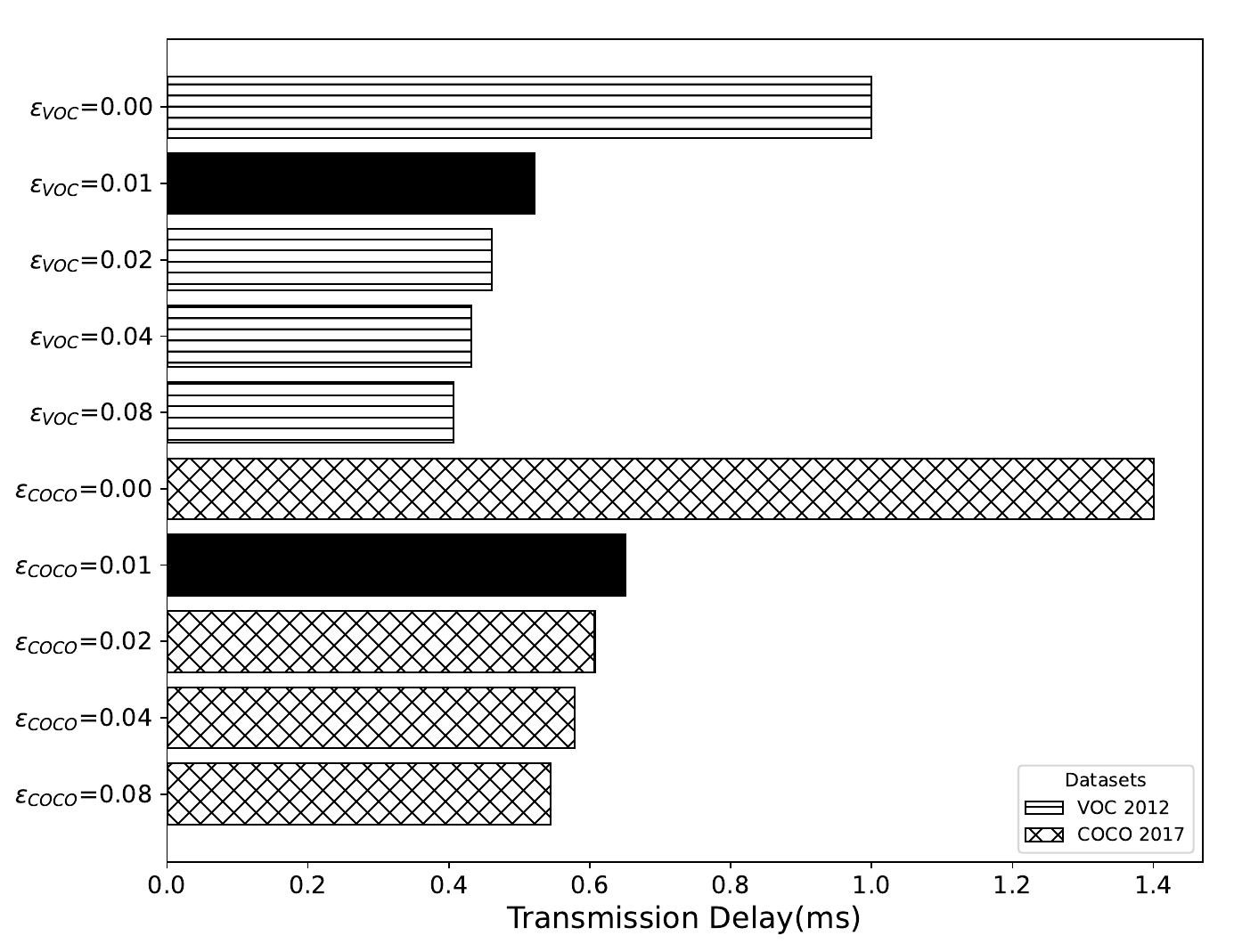} 
    \caption{Effect of $\epsilon$ change on transmission delay for classification tasks.} 
    \label{epsilon-seg}
    \vspace{-5mm}
\end{figure}

\begin{figure}[t]
    \centering
    \includegraphics[width=0.7\linewidth]{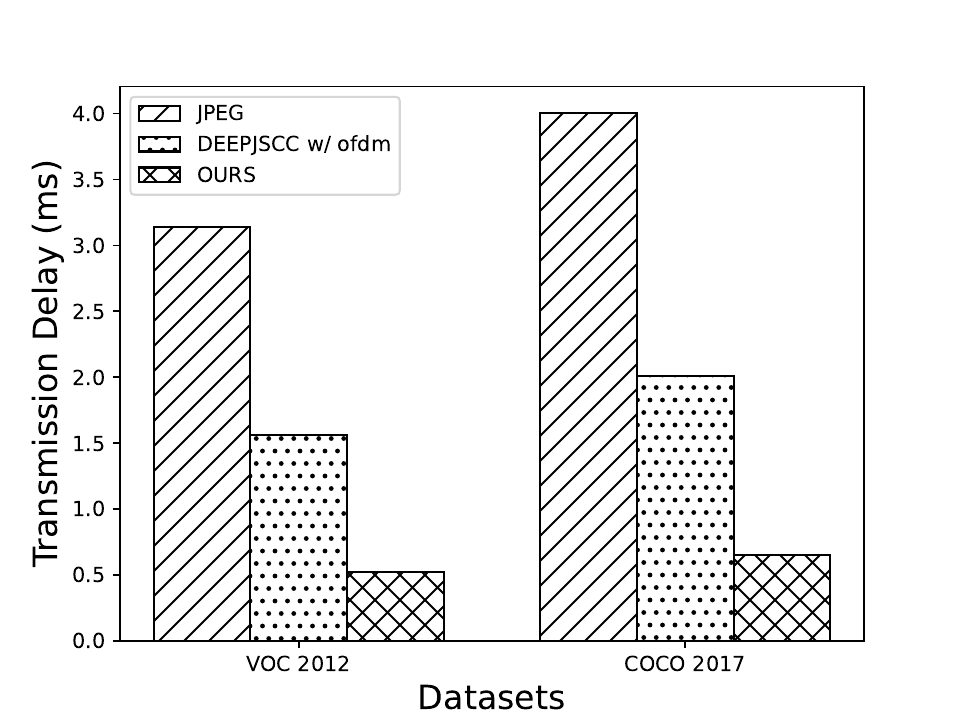} 
    \caption{Transmission latency for classification taks.} 
    \label{tran_delay}
    \vspace{-5mm}
\end{figure}

\begin{figure*}[t]
\centering

\subfloat[]{
\includegraphics[width=0.28\textwidth]{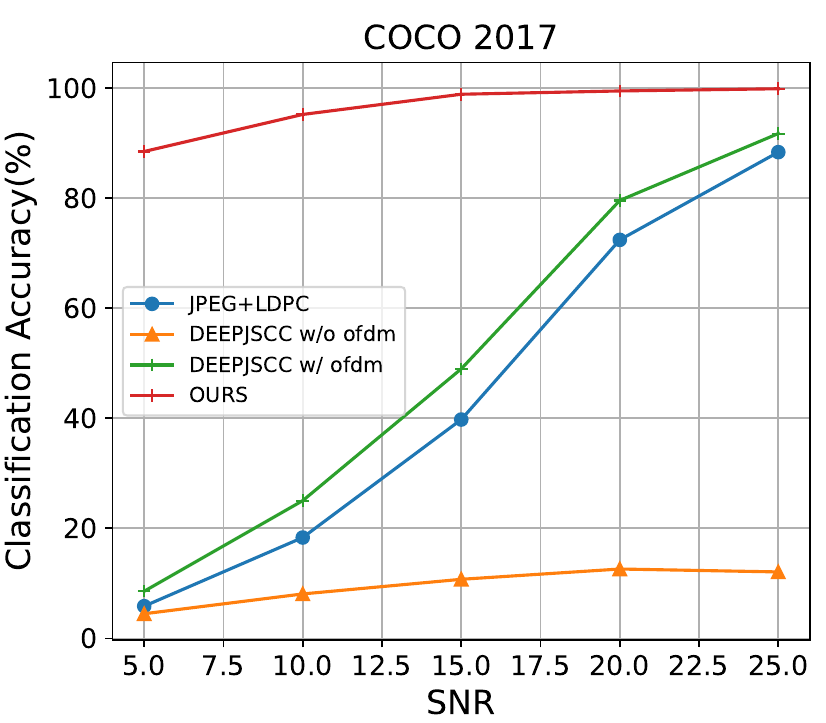}
\label{fig:seg_coco}
}
\subfloat[]{
\includegraphics[width=0.28\textwidth]{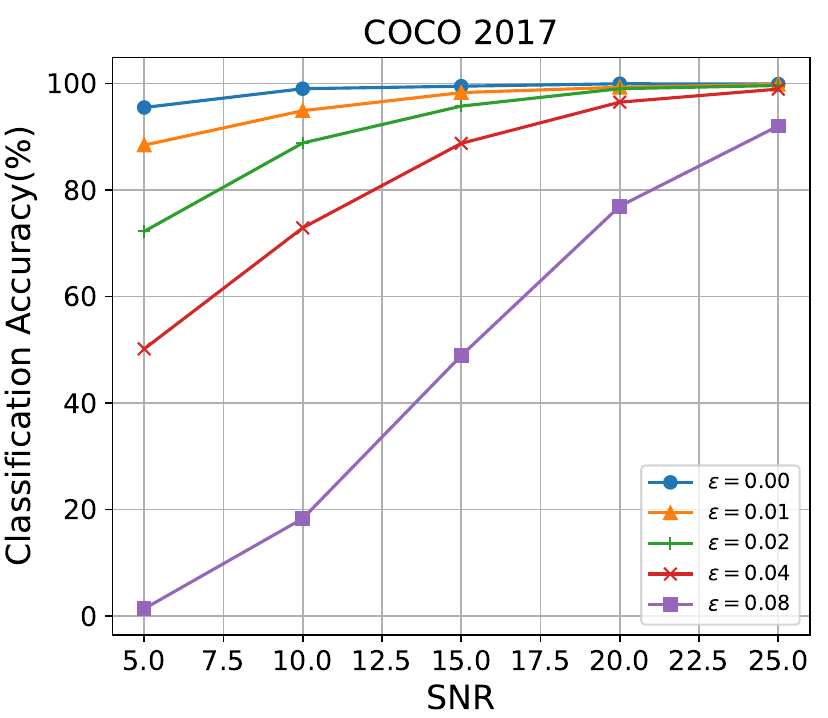}
\label{seg_coco_epsilon}
}
\subfloat[]{
\includegraphics[width=0.28\textwidth]{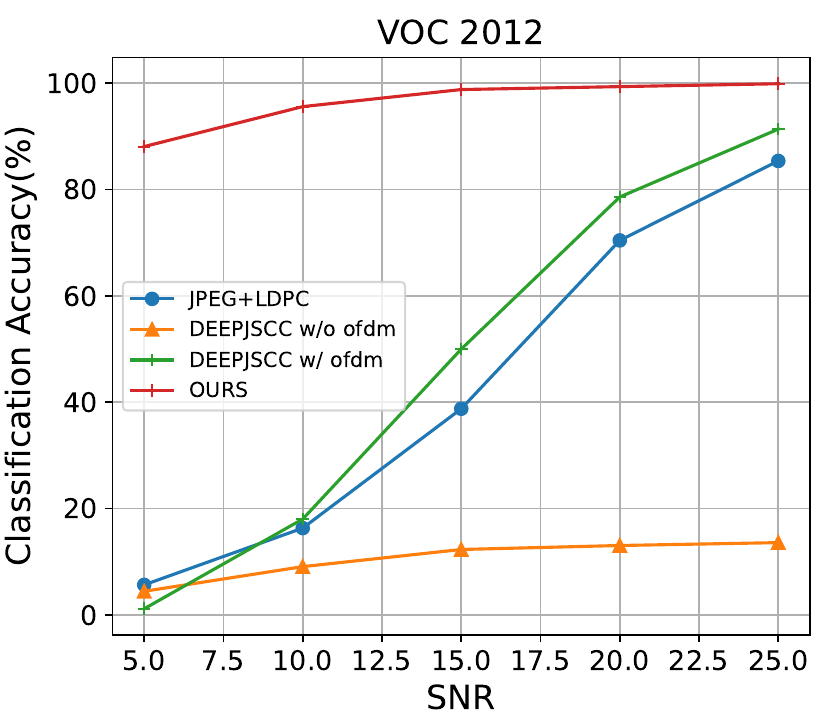}
\label{seg_CA}
}

\vspace{-3mm}

\subfloat[]{
\includegraphics[width=0.28\textwidth]{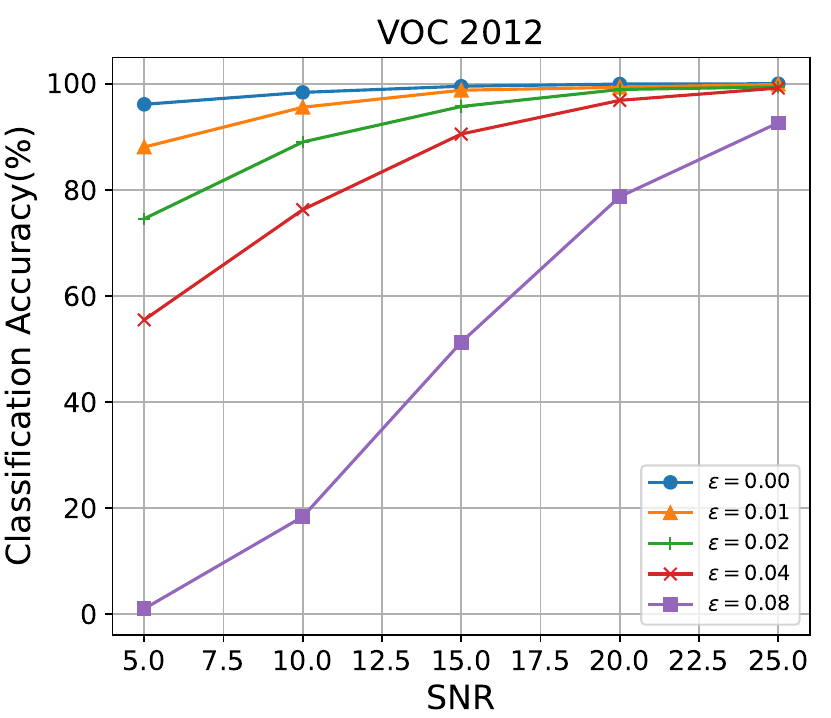}
\label{seg_CA_epsilon}
}
\subfloat[]{
\includegraphics[width=0.28\textwidth]{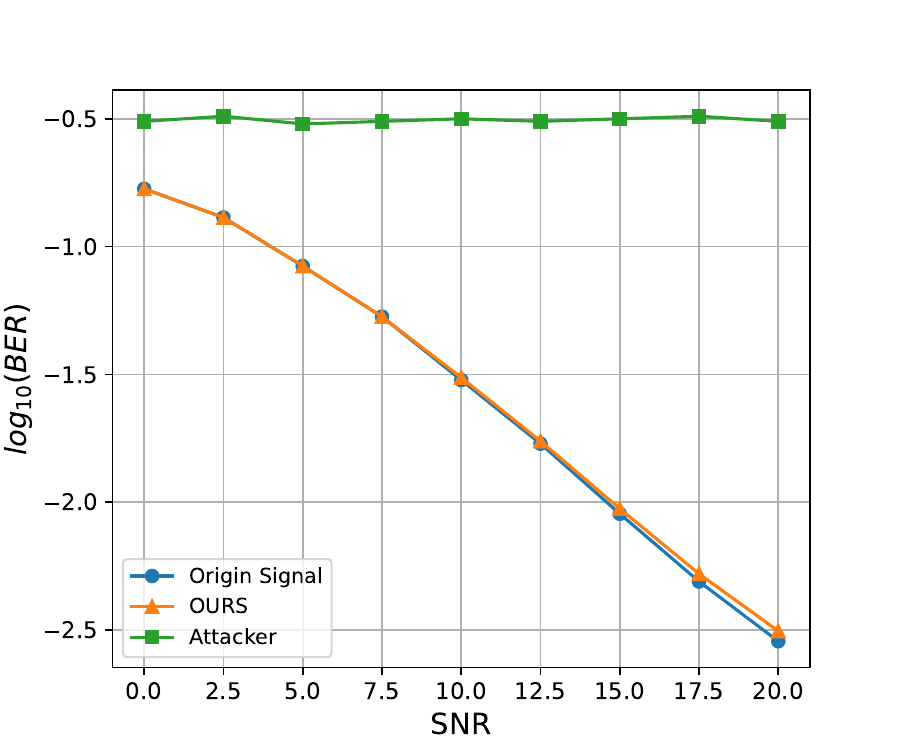}
\label{BER}
}
\subfloat[]{
\includegraphics[width=0.305\textwidth]{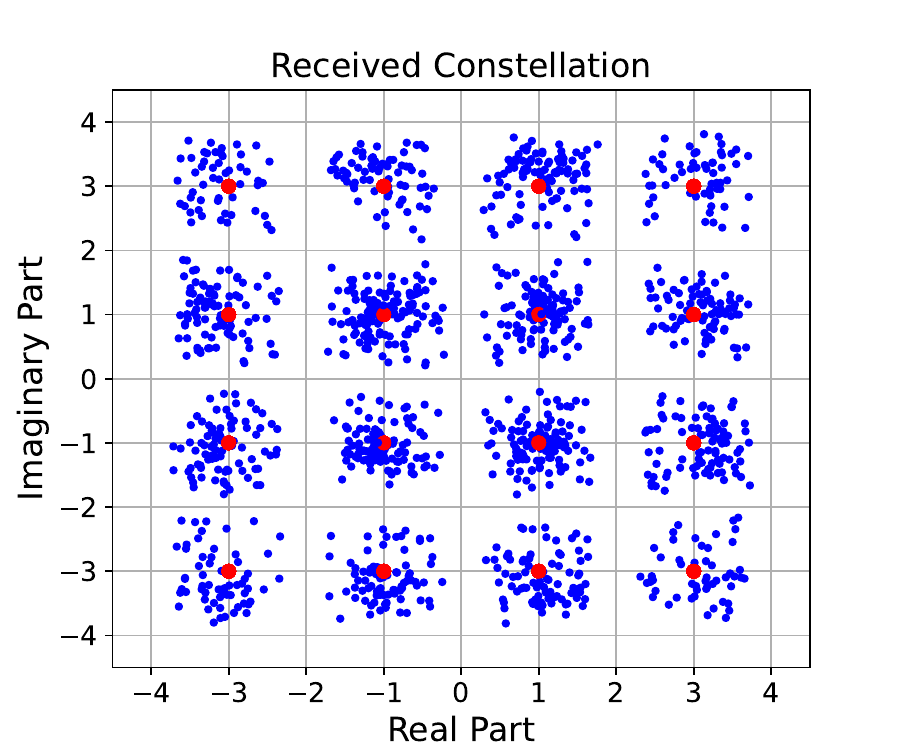}
\label{conmap}
}

\caption{(a)-(d): Performance comparison under different SNRs and datasets for classification tasks. (e): BER performance. (f): Constellation. }
\label{12}
\vspace{-2mm}
\end{figure*}

Fig. \ref{fig:psnr-ssim} shows the reconstructed image performance under different SNRs. It can be seen that the VAE scheme exhibits the worst performance and cannot provide sufficient transmission quality at low SNR, resulting in almost failed transmission. As shown in Fig. \ref{kodak}, due to fully exploring the semantic characteristics of the data, our proposed SemEntropy system exhibits similar performance to state-of-the-art DeepJSCC with OFDM on different datasets and SNRs, even when transmitting 60$\%$ less data. This fully demonstrates the superiority of SemEntropy.

Fig. \ref{casestudy} shows a visual example of the SemEntropy and baseline schemes at an SNR of 20. Under the same conditions, JPEG+LDPC and DeepJSCC exhibit significant detail distortion, and the patterns, colors, and motorcycle details on the helmet are almost indistinguishable. The DeepJSCC scheme combined with OFDM is similar to SemEntropy, but careful observation of helmet and motorcycle details shows that SemEntropy has clearer images and more realistic textures, indicating that traditional image metrics cannot fully reflect the performance of reconstruction. Our scheme has better results in detail processing.

\subsection{Experiments for Classification Tasks}
\subsubsection{Evaluation Metrics and Baselines}
Consider a machine communication scenario that does not require high fidelity data transmission in human communication scenarios, only specific tasks (such as classification tasks) need to be completed. We use classification accuracy and time efficiency as evaluation indicators. In terms of baselines, DeepJSCC with/without OFDM and traditional JPEG+LDPC were used as controls.
\begin{itemize}
    \item \textbf{Accuracy}: The accuracy directly reflects the task completion rate of the semantic communication system, and the formula for accuracy is:
    \[Acc = \frac{T_{\text{correct}}}{T_{\text{total}}},\]
    where \(T_{\text{correct}}\) denotes the count of correctly classified instances, and \(T_{\text{total}}\) signifies the total number of instances.
\end{itemize}

We first conduct experiments on various datasets by employing different values of $\epsilon$. Fig. \ref{epsilon-seg} shows the transmission delay changes when applying SemEntropy on multiple datasets. These datasets cover multiple types of tasks and application scenarios, providing a comprehensive performance evaluation. The results show that as the value $\epsilon$ gradually increases, the transmission delay significantly decreases. When $\epsilon$ is set to 0.01, the transmission delay is significantly reduced, so we chose $\epsilon = 0.01$ as the initial condition in subsequent experiments. Fig. \ref{tran_delay} compares the complexity of SemEntropy and baseline schemes from the perspective of system runtime. The transmission delay is only $25\%$ of the baseline schemes. This reduction not only means more efficient data transmission, but may also lead to a significant reduction in energy consumption, which is particularly important for energy constrained devices such as mobile devices and remote sensors. In addition, this discovery also indicates that under some extreme conditions, even at high $\epsilon$ under certain values, the SemEntropy scheme can still maintain an acceptable level of performance, providing a possibility for effective data processing under strict bandwidth or storage limitations.

Fig. \ref{fig:seg_coco} and Fig. \ref{seg_CA} compare the classification accuracy of SemEntropy and baseline schemes under different SNRs. As shown in Fig. \ref{12} compared to the baseline scheme, the SemEntropy scheme can always achieve better performance. 
SemEntropy maintains a high accuracy gain even at low SNRs, as it selects the most relevant semantic information through semantic entropy constraints.

Fig. \ref{seg_CA_epsilon} and Fig. \ref{seg_coco_epsilon} show in detail the different $\epsilon$ The change in classification accuracy under the epsilon value. Here, $\epsilon$ Represents the amount of relevant semantic information selected under semantic entropy constraints. Larger $\epsilon$ The value means that less semantic information is selected, resulting in a more extreme degree of data compression. The results in Fig. \ref{epsilon-seg} reveal a key finding: when set $\epsilon$ to 0.01, although the SNR fluctuates, the classification accuracy is almost unaffected. This observation indicates that SemEntropy technology can effectively reduce redundant information without sacrificing core task performance, which is particularly important for resource constrained application scenarios.

\subsection{Security of The Proposed Framework}
We utilize the search space for brute-force attacks against the SemEntropy system to quantify the computational cost for potential eavesdroppers. This metric is widely employed in the security analysis of higher-level encryption schemes.

\textbf{The Randomness of Our SKey:}
We generate the semantic key $SKey$ by exploring semantic scores. When attacking the key generator, the attacker attempts to exhaust the search key space through brute force cracking. Search space is expressed: 
\begin{equation}
ss_{scores}=(2^{L_{scores}})^{N}
\end{equation}
where L$_{scores}$ indicates the number of bits per weight and N represents the number of keys in the key stream. The search space for the SKey $ss_{S\!K\!ey}$ can be derived by:
\begin{equation}
ss_{S\!K\!ey}=(2^{L_{S\!K\!ey}})^{N}
\end{equation}
where $L_{S\!K\!ey}$ refers to the length of $SKey$. Therefore, we can conclude that the search space $ss_{S\!K\!ey}$ is much larger than the search space of generating semantic key under a single score, inducing more randomness to secure the semantic communications.

\textbf{The Search Space of Our Key Generator:}
The search space is not only a parameter describing the computational complexity of brute force attacks but also relates to the computational cost of each attack round. The PLK scheme provides protection to the data across multiple modulation stages, posing a decryption challenge to eavesdroppers after executing a portion of the receiving process. If the training symbols remain unchanged, an eavesdropper can detect the signal and store only one complete receiving waveform, subsequently executing a partial reception process to find the correct key. However, in conventional XOR encryption, the XOR operation between the source data and the key sequence occurs in the early stage of physical layer modulation, without randomizing the waveform, enabling the eavesdropper to perform the entire receiving process. In contrast, in our approach, due to the reshuffling of training symbols, the eavesdropper encounters issues in synchronization and channel estimation. This necessitates him/her to store all received signals and redo the entire reception process for each possible key attempt, significantly increasing their computational load. Nevertheless, it's challenging to quantify the number of calculations required for each attack round. Hence, we continue to use the search space to assess the security level of the system since it offers a quantitative description of the required number of attempts.

To attack our key generator, an eavesdropper first needs to enumerate all possible keys in $L_{seedkey}$ bits. The corresponding search space is given as follows.
\begin{equation}
    ss_{seedkey} = 2^{L_{seedkey}}
\end{equation}
In our system, the seed key stream is generated under the constraint of semantic entropy. Unlike previous schemes, due to the dynamic nature of semantic entropy and the randomness of adaptive subcarrier allocation, attackers not only need to attempt to acquire the entire key stream but also to determine the corresponding subcarrier allocation scheme. The total search space can be represented as:
\begin{equation}
    ss_{seedkey}^{\prime} = N \times {2^{({L_{seedkey}})}}^{N}
\end{equation}
Compared to the previous scheme\cite{pre-OFDM}, our method demonstrates an exponential improvement, resulting in a substantially larger search space.

To accurately assess the performance of the proposed digital signal transmission scheme, Bit Error Rate (BER) has been chosen as the primary performance metric. BER is defined as the ratio of incorrectly transmitted bits to the total number of bits sent during the signal transmission process. Fig. \ref{BER} presents a comparison of BER curves for various signals at different levels of SNRs, including the untreated original signal, the signal decrypted by authorized users using the method proposed in this study, and the signal attempted to be decrypted by eavesdroppers. A comparison of the curves in Fig. \ref{BER} reveals that the BER for signals received by authorized users is extremely close to that of the original signal, demonstrating a high degree of consistency. This indicates that the method introduced in this paper provides an effective decryption mechanism for authorized users without compromising signal quality. In contrast, the graph also clearly shows a significant increase in BER for eavesdroppers attempting decryption, approaching a value of 0.5, suggesting an almost complete inability to discern the informational bits within the signal. This result strongly suggests that eavesdroppers are unlikely to obtain any meaningful information from our DLSC system, proving the effectiveness of the proposed method in protecting the security of semantic information.

Furthermore, this study also utilized the 16-QAM OFDM constellation diagram to verify the correctness of OFDM modulation and demodulation during the reception process. As shown in Fig. \ref{conmap}, the encrypted and obfuscated signals can be correctly demodulated, ensuring the accurate recovery of semantic data. This further confirms that our method maintains high efficiency even in complex modulation scenarios, allowing authorized users to transmit and receive signals without hindrance while effectively preventing the leakage of information to unauthorized third parties.

\textbf{Key Generation in Static Environments}
In the context of physical layer key generation, wireless channels are inherently dynamic, offering ample randomness. However, the key generation rate might plummet to near-zero levels in static fading environments. By integrating our $SKey$ with the conventional $PLK$, we not only harness the environment's randomness for semantic encryption and obfuscation but also achieve an expansive search space, denoted as $ss_{total}$. This results in a significantly enhanced security level for the system.

\section{CONCLUSION}
In this work, we propose a novel semantic communication framework, SemEntropy, aiming at jointly optimizing transmission efficiency and security by measuring semantic importance to guide adaptive subcarrier allocation and secure communication with physical layer encryption. Specifically, we use a carefully designed semantic information generator to extract semantic information more accurately through importance measurement and use semantic entropy to select the transmitted semantic information, thereby achieving efficient semantic transmission. At the same time, through the study of semantic derivation, the physical layer semantic key was used to encrypt the system in a static fading wireless environment, and quantitative analysis was conducted, indicating that the search space of this method is much larger than previous work. Experiments show that the transmission performance of SemEntropy is superior to traditional communication systems and existing semantic communication systems, especially in low SNR environments. In the future, we will deploy the proposed SemEntropy to a real radio system.

\end{document}